\documentclass[graybox]{svmult}

\usepackage{type1cm}        
\usepackage{makeidx}         
\usepackage{graphicx}        
\usepackage{multicol}        
\usepackage[bottom]{footmisc}

\usepackage{newtxtext}       %
\usepackage{newtxmath}       

\begin{document}
\setcounter{chapter}{12}

\title{Testing Gravity and Predictions Beyond the Standard Model at Short
Distances: The Casimir Effect}
\titlerunning{Testing Gravity and Predictions Beyond the Standard Model}
\author{Galina L. Klimchitskaya and Vladimir M. Mostepanenko}
\institute{Galina L. Klimchitskaya \at Central Astronomical Observatory at Pulkovo
of the Russian Academy of Sciences, Saint Petersburg, 196140, Russia;
Peter the Great Saint Petersburg
Polytechnic University, Saint Petersburg, 195251, Russia,
\email{g.klimchitskaya@gmail.com}
\and Vladimir M. Mostepanenko \at Central Astronomical Observatory at Pulkovo
of the Russian Academy of Sciences, Saint Petersburg, 196140, Russia;
Peter the Great Saint Petersburg
Polytechnic University, Saint Petersburg, 195251, Russia;
Kazan Federal University, Kazan, 420008, Russia,
\email{vmostepa@gmail.com}}
%
%
\maketitle
\label{Chap:0204_CE}

\setcounter{minitocdepth}{3}

\dominitoc

\abstract{The Standard Model of elementary particles and their interactions
does not include the gravitational interaction and faces problems in
understanding of the dark matter, dark energy, strong CP violation etc.
In continuing attempts to solve these problems, many predictions of new
light elementary particles and hypothetical interactions beyond the
Standard Model have been made. These predictions can be constrained by
many means and, specifically, by measuring the Casimir force arising
between two closely spaced bodies due to the zero-point and thermal
fluctuations of the electromagnetic field. After a brief survey in the
theory of the Casimir effect, the strongest constraints on the power-type
and Yukawa-type corrections to Newtonian gravity, following from
measuring the Casimir force at short distances, are considered. Next,
the problems of dark matter, dark energy and their probable constituents
are discussed. This is followed by an analysis of constraints on the
dark matter particles, and, specifically, on axions and axionlike
particles, obtained from the Casimir effect. The question of whether the
Casimir effect can be used for constraining the spin-dependent interactions
is also considered. Then the constraints on the dark energy particles,
like chameleons and symmetrons, are examined. In all cases the subject
of our treatment is not only measurements of the Casimir force but some
other relevant table-top experiments as well. In conclusion, the prospects
of the Casimir effect for constraining theoretical predictions beyond the
Standard Model at short distances are summarized.}





\newcommand{\kb}{{k_{\bot}}}
\newcommand{\skb}{{k_{\bot}^2}}
\newcommand{\vk}{{\mbox{\boldmath$k$}}}
\newcommand{\rv}{{\mbox{\boldmath$r$}}}
\newcommand{\ve}{{\varepsilon}}
\newcommand{\xk}{{(i\xi_l,k_{\bot})}}

\section{Introduction: Gravity, the Standard Model and Beyond}
\label{secKM:1}

The gravitational force is familiar to everybody from the day-to-day
experience. If some body is released, it falls to earth under the
influence of gravitational attraction. The laws of free fall were
experimentally discovered by Galileo Galilei who found that in a
vacuum the bodies of different weight fall with a uniform acceleration
and reach the earth concurrently. This great (and somewhat counterintuitive)
result was later derived theoretically by Newton from his second law and
law of gravity under a fundamental assumption that the inertial and
gravitational masses are equal (the equivalence principle).

It is common knowledge that according to Newton's law of gravity two
point masses $m_1$ and $m_2$ separated by a distance $r$ attract each other
with the force
\begin{equation}
F_{gr}(r)=-\frac{dV_{gr}(r)}{dr}=-G\frac{m_1m_2}{r^2},
\label{eqKM1}
\end{equation}
where $G$ is the gravitational constant and $V_{gr}$ is the gravitational
interaction energy
\begin{equation}
V_{gr}(r)=-G\frac{m_1m_2}{r}.
\label{eqKM2}
\end{equation}

Einstein's general relativity theory \cite{KM1} changed seriously the
conceptual pattern of gravity. According to this theory, gravity is a
curved space-time whose geometrical properties are determined not only
by the masses of material bodies but by all components of their
stress-energy tensor. It is important, however, that corrections to
(\ref{eqKM1}) and (\ref{eqKM2}) predicted by the general relativity
theory for mass and separation scales characteristic of a physical
laboratory are negligibly small \cite{KM2}. The gravitational force
ensures the stability of planets, solar system, galaxies, and
determines the structure and evolution of the whole Universe.

Another force which manifests itself in day-to-day life is the
electromagnetic one. The classical theory of this force was created by
Maxwell and is known as classical electrodynamics \cite{KM3}. Unlike
the gravitational force which is universal and acts between all
material bodies, the electromagnetic force acts only between bodies
possessing electric charges. The Maxwell theory describes only the
classical aspects of electromagnetic interaction, whereas the full
picture is given by quantum electrodynamics \cite{KM4} created in the
middle of the last century by Feynman, Schwinger, Tonomaga, and Dyson.
Electromagnetic forces bind nuclei and electrons into atoms, create
chemical bonds which make possible the existence of molecules. They are
responsible for the structure of crystal lattices and are heavily used
in electronics and all modern technologies.

Two other types of fundamental forces existing in nature, weak and
strong interaction, are entirely quantum. They are not visible to the
naked eye. The weak interaction is responsible for a decay of many
elementary particles whereas the strong interaction binds protons and
neutrons into atomic nuclei. In the middle of sixties of the last
century, Weinberg, Salam and Glashow developed the unified theory of
weak and electromagnetic interactions \cite{KM5}. For electromagnetic
interaction, the intermediate particles between electrically charged
particles (the so-called force carriers) are the massless photons. Photons
are the specific case of gauge bosons, i.e., particles of spin one
which mediate different interactions. The force carriers for a weak
interaction between particles are the three massive bosons, two of
which, $W^+$ and $W^-$, are electrically charged and one, $Z_0$, is
neutral.

By the middle of seventies, owing to works by Nambu, Gross, Wilczek,
Politzer and other scientists, the theory of strong interactions
had been elaborated. According to this theory, strongly interacting
particles, e.g., protons and neutrons, consist of quarks possessing
spin 1/2 and the new type of charge called color. Unlike the electric
charge, which may be either positive or negative, the color charge has
three different values (and respective anticolors). The force carriers
for a strong interaction between quarks are eight massless gauge
bosons called gluons which bear the color charges. Due to this the
theory of strong interactions was called quantum chromodynamics~\cite{KM6}.

The Standard Model is a unified theory of the three fundamental
interactions -- electromagnetic, weak, and strong \cite{KM7}.
According to the Standard Model, there are three pairs (generations)
of quarks possessing the color charge and three pairs of spin 1/2
particles called leptons (the most familiar of them are electrons
and respective electronic neutrinos). There are also as many antiquarks and
antileptons as quarks and leptons. Next, the Standard Model includes
the force carriers of electromagnetic, weak and strong interactions,
i.e., photons, three massive bosons, $W^+, W^-, Z^0$, and eight gluons.
Finally, an important element of the Standard Model is the heavy
particle of zero spin called the Higgs boson predicted by Higgs in
1964, which is responsible for a generation of masses of other
elementary particles. By now all the above elements of the Standard
Model were observed in the accelerator experiments and many other
theoretical predictions made on this basis found their experimental
confirmation.

Great successes of the Standard Model in particle physics do not
mean, however, that we already have the theory of everything. The
major problem is that the general relativity theory remains to be
isolated from the Standard Model, and the gravitational force avoids
unification with other three fundamental interactions in spite of
persistent efforts undertaken during several decades. Both Newtonian
gravity and Einstein's general relativity are the entirely classical
theories. However, for a description of physical phenomena happening
in the close proximity of space-time singularities predicted by the
general relativity theory, one evidently needs some theory which takes
into account the quantum effects. There are also unresolved problems of
dark matter and dark energy which are observed only indirectly through
their gravitational interactions but are not explained in the context
of the Standard Model. It should be mentioned also that at short
distances below a micrometer the Newton law (\ref{eqKM1}) lacks of
experimental confirmation and leaves a room for modifications at
the cost of different quantum effects.

There are also other serious problems of the Standard Model. Among
them one should mention the hierarchy problem, i.e., an unanswered
question of why there is a difference by the factor of $10^{24}$
between the strength of weak and gravitational interactions, the
problem of neutrino mass, which was zero in the original formulation
of the model but turns out to be nonzero according to precise
measurements, and the problem of an asymmetry between matter and
antimatter. An important problem is also the strong CP violation
(i.e., the violation of invariance relative to the charge conjugation
accompanied by the parity transformation) which is admitted by the
formalism of quantum chromodynamics but is not observed in experiments
involving only the strong interaction.

All these problems are widely discussed in the literature, and many
theoretical approaches to their resolution are proposed in the
framework of extended standard model, supersymmetry, supergravity
\cite{KM8}, and string theory \cite{KM9} {(see also the discussions in part \ref{PART1} of this book)}. 
The mentioned approaches go
beyond the Standard Model and introduce additional particles,
interactions, and symmetries leading to some theoretical predictions
which can be verified experimentally using the powerful high energy
accelerators, astrophysical observations, and laboratory experiments.

Below we consider some of these predictions which can be verified in
experiments on measuring the Casimir force arising between two
closely spaced uncharged material bodies due to the zero-point and
thermal fluctuations of the electromagnetic field. As it is shown
below, these relatively cheap and compact laboratory experiments can
compete with huge accelerators in testing some important theoretical
predictions beyond the Standard Model.

\section{Electromagnetic Casimir Force and the Quantum Vacuum}
\label{secKM:2}

In 1948, Casimir \cite{KM10} considered two parallel, uncharged ideal metal planes in
vacuum at zero temperature spaced at a distance $a$ and calculated the zero-point energy
of the electromagnetic field in the presence and in the absence of these planes, i.e.,
in free space. The case with the planes differs in that the tangential component of
electric field and the normal component of magnetic induction vanish on their surfaces.
Casimir considered a difference between the zero-point energies per unit area
in the presence and in the absence of  planes
\begin{equation}
E(a)=\hbar\int_0^{\infty}\frac{\kb d\kb}{2\pi}\left(
\sum_{l=0}^{\infty}{\vphantom{\sum}}^{\prime}\omega_{\kb,l}-
\frac{a}{\pi}\int_0^{\infty}dk_z\omega_k\right),
\label{eqKM3}
\end{equation}
\noindent
where $\vk=(k_x,k_y,k_z)$ is the wave vector, $\kb=\sqrt{k_x^2+k_y^2}$ is the magnitude
of the wave vector projection on the planes, the prime on the summation sign
divides the term with $l=0$ by 2, and the frequencies of the zero-point oscillations are
given by the following expressions:
\begin{equation}
\omega_{\kb,l}=c\sqrt{\skb+\left(\frac{\pi l}{a}\right)^2}, \qquad
\omega_k=c\sqrt{\skb+k_z^2}.
\label{eqKM4}
\end{equation}

Although both terms on the right-hand side of (\ref{eqKM3}) are infinitely large, their
difference is finite. Using the Abel-Plana formula for a difference between the sum and
the integral \cite{KM11}, one obtains
\begin{equation}
E(a)=-\frac{\pi^2\hbar c}{a^3}\int_0^{\infty}\!\!\!\!ydy
\int_y^{\infty}\!\frac{\sqrt{t^2-y^2}}{e^{2\pi t}-1}dt.
\label{eqKM5}
\end{equation}

Then, calculating the integrals in  (\ref{eqKM5}), one arrives at the famous Casimir result
\begin{equation}
E(a)=-\frac{\pi^2}{720}\,\frac{\hbar c}{a^3}
\label{eqKM6}
\end{equation}
\noindent
and at respective expression for the Casimir force per unit area of the plates
\begin{equation}
P(a)=-\frac{d E(a)}{d a}=-\frac{\pi^2}{240}\,\frac{\hbar c}{a^4},
\label{eqKM7}
\end{equation}
\noindent
i.e., the Casimir pressure. This pressure is some kind of a macroscopic quantum effect
determined entirely by the zero-point oscillations of quantized electromagnetic field.
Thus, for two ideal metal planes separated by a distance $a=1~\mu$m we obtain from
(\ref{eqKM7}) an attractive pressure $P(a)=-1.3~$mPa.

In a physical laboratory  we deal not with ideal metals but with real material bodies
made of metallic, dielectric or semiconductor materials. In 1955, Lifshitz \cite{KM12}
created the general theory describing the free energy and force arising between two thick
material plates (semispaces) spaced at a separation $a$ in thermal equilibrium with the
environment at temperature $T$. The material properties in this theory were characterized
by the dielectric permittivities $\ve^{(n)}(\omega)$ of the first and second plates ($n=1,\,2$).
Later the Lifshitz results were generalized for the plates possessing magnetic properties
characterized by the magnetic permeabilities  $\mu^{(n)}(\omega)$ \cite{KM13}.

In the framework of the Lifshitz theory, the free energy of interaction caused by the
zero-point and thermal fluctuations of the electromagnetic field per unit area of the
plates is given by \cite{KM12,KM14}
\begin{equation}
{\cal F}(a,T)=\frac{k_BT}{2\pi}\sum_{l=0}^{\infty}{\vphantom{\sum}}^{\prime}
\int_0^{\infty}\kb d\kb\sum_{\alpha}\ln\left[1-r_{\alpha}^{(1)}\xk r_{\alpha}^{(2)}\xk
e^{-2aq_l}\right],
\label{eqKM8}
\end{equation}
\noindent
where $k_B$ is the Boltzmann constant, $\xi_l=2\pi k_BTl/\hbar$ are the Matsubara frequencies,
$q_l=\sqrt{\skb+\xi_l^2/c^2}$, and the reflection coefficients for two independent
polarizations of the electromagnetic field, transverse magnetic ($\alpha={\rm TM}$) and
transverse electric ($\alpha={\rm TE}$), are given by
\begin{eqnarray}
&&
r_{\rm TM}^{(n)}\xk=
\frac{\ve^{(n)}(i\xi_l)q_l-k^{(n)}\xk}{\ve^{(n)}(i\xi_l)q_l+k^{(n)}\xk},
\nonumber \\
&&
r_{\rm TE}^{(n)}\xk=
\frac{\mu^{(n)}(i\xi_l)q_l-k^{(n)}\xk}{\mu^{(n)}(i\xi_l)q_l+k^{(n)}\xk},
\label{eqKM9}
\end{eqnarray}
\noindent
where
\begin{equation}
k^{(n)}\xk=\sqrt{\skb+\ve^{(n)}(i\xi_l)\mu^{(n)}(i\xi_l)\frac{\xi_l^2}{c^2}}.
\label{eqKM10}
\end{equation}

In a similar way, the Casimir force per unit area of real material plates is expressed as
\begin{eqnarray}
&&
P(a,T)=-\frac{\partial{\cal F}(a,T)}{\partial a}=
-\frac{k_BT}{\pi}\sum_{l=0}^{\infty}{\vphantom{\sum}}^{\prime}
\int_0^{\infty}q_l\kb d\kb
\nonumber \\
&&~~~~~~~~~~~\times
\sum_{\alpha}\left[\frac{e^{2aq_l}}{r_{\alpha}^{(1)}\xk r_{\alpha}^{(2)}\xk}-1
\right]^{-1}\!\!\!.
\label{eqKM11}
\end{eqnarray}

Taking into account that ideal metal is the perfect reflector, so that
\begin{equation}
r_{\rm TM}^{(n)}\xk=-r_{\rm TE}^{(n)}\xk=1
\label{eqKM12}
\end{equation}
\noindent
at all $\xi_l$, at $T=0$ one obtains from (\ref{eqKM8}) and (\ref{eqKM11}) the Casimir
results (\ref{eqKM6}) nd (\ref{eqKM7}). In the limiting case of small separations,
(\ref{eqKM8}) and (\ref{eqKM11}) describe the familiar van der Waals force which depends
on $\hbar$ but does not depend on the speed of light $c$. In the opposite limiting case
of large separations, the resulting free energy and force do not depend either on $\hbar$
or $c$. This is the so-called classical regime where the Casimir interaction depends
only on $T$.

Precise measurements of the Casimir force allowing quantitative comparison between experiment
and theory were performed by means of an atomic force microscope, whose sharp tip was replaced
with a relatively large sphere, and a micromechanical torsional oscillator (see \cite{KM14,KM15}
for a review). All these experiments measured the Casimir force not between two parallel
plates but between a sphere and a plate. The Casimir force between a sphere of radius $R$
and a plate $F^{SP}(a,T)$ can be calculated in the framework of the Lifshitz theory using
the proximity force approximation  \cite{KM14,KM15}
\begin{equation}
F^{SP}(a,T)=2\pi R{\cal F}(a,T),
\label{eqKM13}
\end{equation}
\noindent
where the Casimir free energy between two parallel plates ${\cal F}$
is given by the Lifshitz formula
(\ref{eqKM8}). Exact calculations of the Casimir force in sphere-plate geometry using the
scattering approach \cite{KM16,KM17,KM18,KM19} and the gradient expansion
\cite{KM20,KM21,KM22,KM23,KM24} have shown that the errors introduced by (\ref{eqKM13})
are less than $a/R$, i.e., less than a fraction of a percent in the most of experimental
configurations.

By calculating the derivative of (\ref{eqKM13}) with respect to separation, one can
express another quantity measured in many experiments, i.e., the gradient of the Casimir
force in sphere-plate geometry via the Casimir force (\ref{eqKM11}) per unit area of two
parallel plates
\begin{equation}
\frac{\partial}{\partial a}F^{SP}(a,T)=-2\pi R P(a,T).
\label{eqKM14}
\end{equation}

For comparison of theoretical predictions with the measurement data of precise experiments,
one should compute the Casimir free energy (\ref{eqKM8}) and the Casimir pressure (\ref{eqKM11})
with sufficient precision. To do so, one needs to have the values of dielectric permittivities
of plate materials at sufficiently large number of pure imaginary Matsubara frequencies.
This is usually achieved by means of the Kramers-Kronig relation using the measured optical
data for the complex indices of refraction of plate materials. In doing so the terms of the
Lifshitz formulas (\ref{eqKM8}) and (\ref{eqKM11}) with $l=0$ play an important role in
obtaining the physically correct results.

Unfortunately, the optical data are available at only sufficiently high frequencies
$\omega\geqslant\omega_{\min}$. Because of this, the obtained dielectric permittivity
is usually extrapolated down to zero frequency using some theoretical model.
For experiments with metallic test bodies, which are used below for testing the predictions
beyond the Standard Model, the most reasonable extrapolation seems to be by means of the
well tested Drude model. In this case, the dielectric permittivities of plate materials
take the form
\begin{equation}
\ve_D^{(n)}(i\xi_l)=\ve_c^{(n)}(i\xi_l)+\frac{\omega_{p,n}^2}{\xi_l(\xi_l+\gamma_n)},
\label{eqKM15}
\end{equation}
\noindent
where $\ve_c^{(n)}(i\xi_l)$ is a contribution due to core electrons determined by the
optical data, $\omega_{p,n}$ is the plasma frequency and $\gamma_n$ is the relaxation
parameter.

It turned out, however, that the measurement data of all precise experiments with
nonmagnetic (Au) metals \cite{KM25,KM26,KM27,KM28,KM29,KM30,KM31,KM32,KM33} and
magnetic (Ni) metals \cite{KM34,KM35,KM36,KM37} exclude the theoretical predictions of
the Lifshitz theory using the dielectric functions (\ref{eqKM15}). Specifically, for two
test bodies made of Ni a disagreement between experiment and theory in measurements of
the differential Casimir force is up to a factor of 1000 \cite{KM37}.
If, however, one makes an extrapolation by means of the plasma model, i.e.,
puts $\gamma_n=0$ in (\ref{eqKM15}),
\begin{equation}
\ve_p^{(n)}(i\xi_l)=\ve_c^{(n)}(i\xi_l)+\frac{\omega_{p,n}^2}{\xi_l^2},
\label{eqKM16}
\end{equation}
\noindent
the predictions of the Lifshitz theory come to a very good agreement with the
measurement data of all precise experiments
\cite{KM25,KM26,KM27,KM28,KM29,KM30,KM31,KM32,KM33,KM34,KM35,KM36,KM37}.

This situations calls for some clarification because at low frequencies conduction
electrons really possess relaxation properties described by the phenomenological
parameter $\gamma_n$. It is then unclear why one should put  $\gamma_n=0$ in computations
of the Casimir force. Although the ultimate answer to this question is not found yet,
theory suggests some plausible explanation. First of all, it was proven
\cite{KM38,KM39,KM40,KM41} that for metals with perfect crystal lattices the Casimir
entropy calculated using the dielectric permittivity (\ref{eqKM15}) violates the
third law of thermodynamics, the Nernst heat theorem, but satisfies it if the permittivity
(\ref{eqKM16}) is used.

Next, for graphene, which is a novel 2D material \cite{KM42}, the dielectric permittivity
is not of a model character. At low energies characteristic for the Casimir effect,
the dielectric properties of graphene can be calculated on the basis of first principles
of quantum electrodynamics at nonzero temperature using the polarization tensor in
(2+1)-dimensional space-time \cite{KM43,KM44}. It was found that graphene is described
by two spatially nonlocal dielectric permittivities, i.e., depending on both the frequency
$\omega$ and the 2D wave vector $\vk$ \cite{KM45,KM46}. The Lifshitz theory using these
permittivities turned out to be in perfect agreement with measurements of the Casimir
force from graphene \cite{KM47,KM48,KM49,KM50} and with the Nernst heat theorem
\cite{KM51,KM52,KM53,KM54,KM55}.

This suggests that the model dielectric permittivity (\ref{eqKM15}), which is well-checked
for the propagating electromagnetic waves on the mass shell in vacuum,
may be inapplicable to the
evanescent (off-the-mass-shell) waves. The latter contribute essentially to the Casimir
free energy and force (\ref{eqKM8}) and (\ref{eqKM11}) caused by the electromagnetic
fluctuations. First steps on the road to justification of this conjecture were made
by the recently proposed spatially nonlocal dielectric permittivities which describe
nearly the same response, as does the Drude model, to the propagating waves but an
alternative response to the evanescent ones \cite{KM56,KM56a}. The Lifshitz theory employing
these permittivities is in as good agreement with measurements of the Casimir force
between nonmagnetic metals and with the Nernst heat theorem as when it uses the plasma
model (\ref{eqKM16}) \cite{KM56,KM56a,KM57}. Recently it was also shown that it agrees equally
well with measurements of the Casimir force between magnetic metals \cite{KM56a,KM58}.

By and large one can conclude that although there is a continuing discussion in the
literature on theoretical description of the Casimir interaction between real material
bodies (see \cite{KM59} for a review), the predictions of the Lifshitz theory are now
found in good agreement with the measurement data of all precise experiments and the
measure of this agreement can be used for constraining the hypothetical forces of
nonelectromagnetic origin.

\section{Testing the Power-Type Corrections to Newtonian Gravity from the Casimir Effect}
\label{secKM:3}

{}From the point of view of quantum field theory, the gravitational interaction energy
(\ref{eqKM2}) can be considered as originating from an exchange of one massless particle
between two massive particles $m_1$ and $m_2$. Exactly in this way the Coulomb potential
is derived in quantum electrodynamics by considering an exchange of one photon between
two charged particles.

The Standard Model does not contain massless particles in a free state except of photons
(gluons are confined inside of barions). There are, however, massless particles predicted
by some extensions of the Standard Model. For instance, theory of electroweak interactions
with an extended Higgs sector predicts pseudoscalar massless particles called arions
\cite{KM60}.  An exchange of one arion between electrons belonging to atoms of two
neighboring test bodies leads to the spin-dependent effective potential which averages
to zero when integrating over their volumes. The spin-independent effective potential
decreasing with separation as $r^{-3}$ arises from the process of two-arions exchange
\cite{KM61}.

In a similar way, the effective potential decreasing with separation as $r^{-5}$
arises from an exchange of neutrino-antineutrino pair between two neutrons \cite{KM62,KM63}.
The power-type potentials result also from an exchange of even numbers of goldstinos which
are the massless fermions introduced in the theoretical schemes with a spontaneously
broken supersymmetry \cite{KM64} and other predicted particles.

Taking into account that the power-type interactions with different powers coexist with the
gravitational potential, the resulting interaction energy is usually represented as
\begin{equation}
V_l(r)=-\frac{Gm_1m_2}{r}\left[1+\Lambda_l\left(\frac{r_0}{r}\right)^{l-1}\right],
\label{eqKM17}
\end{equation}
\noindent
where $\Lambda_l$ is the dimensionless interaction constant, $l=1,\,2,\,3,\,\ldots$,
and $r_0$ with the dimension of length is introduced to preserve the dimension of
energy for $V_l(r)$. Following many authors, we put  $r_0=1~\mbox{F}=10^{-15}~$m.
For $l=1$, the quantity $1+\Lambda_1$ has the meaning of a factor connecting the values
of inertial and gravitational masses, for $l=3$ the second term in (\ref{eqKM17})
presents a correction to the Newtonian potential due to an exchange of two arions, and
for $l=5$ --- due to an exchange of neutrino-antineutrino pair.

The power-type corrections to Newton's law arise not only due to an exchange of massless
hypothetical particles but in extensions of the Standard Model which exploit
the extra-dimensional
unification schemes with noncompact but warped extra dimensions. In this case, the modified
gravitational interaction energy at separations $r\gg K_w$ takes the form
\cite{KM65,KM66}
\begin{equation}
V_3(r)=-\frac{Gm_1m_2}{r}\left(1+\frac{2}{3K_w^2r^2}\right),
\label{eqKM18}
\end{equation}
\noindent
where $K_w$ is the warping scale. This is the potential of the form of (\ref{eqKM17})
with $\Lambda_3=2/(3K_w^2r_0^2)$.

Constraints on the values of interaction constant $\Lambda_l$ with different $l$ can be
obtained from the gravitational experiments of E\"{o}tvos and Cavendish type.
In the E\"{o}tvos-type experiments one verifies a validity of the equivalence principle,
i.e., places limits on possible deviations between the inertial and gravitational masses.
Using (\ref{eqKM17}), these limits can be recalculated in the constraints on $\Lambda_1$.
Thus, from the most precise short-range E\"{o}tvos-type experiments \cite{KM67,KM68} the
constraint $|\Lambda_1|\leqslant 1\times 10^{-9}$ was obtained.

In the Cavendish-type experiments, one measures probable deviations of the force acting
between two bodies from the Newton law (\ref{eqKM1}). {}From the power-type interaction
energy (\ref{eqKM17}) one finds the respective force
\begin{equation}
F_l(r)=-\frac{d V_l(r)}{d r}=
-\frac{Gm_1m_2}{r^2}\left[1+l\Lambda_l\left(\frac{r_0}{r}\right)^{l-1}\right].
\label{eqKM19}
\end{equation}
\noindent
Then the constraints on $\Lambda_l$ can be found from the measured limits on the
dimensionless quantity
\begin{equation}
\varepsilon_l=\frac{1}{rF_l(r)}\frac{d}{d r}\left[r^2F_l(r)\right],
\label{eqKM20}
\end{equation}
\noindent
which is equal to zero if $\Lambda_l=0$, i.e., no power-type interaction in addition to
gravity is present. Using this approach, from the Cavendish-type experiment \cite{KM69}
the following constraints on $\Lambda_l$ were obtained \cite{KM70}:
$|\Lambda_2|\leqslant 4.5\times 10^{8}$,
$|\Lambda_3|\leqslant 1.3\times 10^{20}$,
$|\Lambda_4|\leqslant 4.9\times 10^{31}$,
$|\Lambda_5|\leqslant 1.5\times 10^{43}$.

In \cite{KM61,KM71} it was suggested to obtain constraints on the power-type interactions
from measurements of the Casimir force. The Casimir force
$F^{LP}(a)$ between a spherical lens
 of centimeter-size radius and a plate  both made of quartz was measured at
distances $a\leqslant 1~\mu$m in \cite{KM72} with a relative error
$\Delta F/F^{LP}\approx 10$\%, where $\Delta F$ is the absolute error.
In the limits of this error, the measurement data were found to be in agreement with
theoretical predictions of the Lifshitz theory.

Any hypothetical interaction energy of power type between an atom of the lens at a point
{\boldmath$r$}$_1$ and an atom of the plate  at a point
{\boldmath$r$}$_2$ is given by (\ref{eqKM17}) where
$r=|{\mbox{\boldmath$r$}}_1-{\mbox{\boldmath$r$}}_2|$.
Then, the total interaction force between the experimental test bodies
(the lens and the plate) is given by the
integration over their volumes $V_1$ and $V_2$
with subsequent negative differentiation with respect to
the distance $a$ of their closest approach
\begin{equation}
F_l^{LP}(a)=-n_1n_2\frac{\partial}{\partial a}\int_{V_1}\!\!\!d^3r_1\int_{V_2}\!\!\!d^3r_2
V_l(|{\mbox{\boldmath$r$}}_1-{\mbox{\boldmath$r$}}_2|),
\label{eqKM21}
\end{equation}
\noindent
where $n_1$ and $n_2$ are the numbers of atoms per unit volume of the first and second
test bodies. In doing so, one can neglect by the Newtonian contribution on the right-hand
side of (\ref{eqKM21}) because it is negligibly small as compared to the experimental
error in the micrometer separation range.

Taking into account that no additional interaction was observed within the limits of
measurement errors, the constraints on $\Lambda_l$ with $l=1$, 2, 3, 4, and 5 were
obtained from the inequality \cite{KM61,KM71}
\begin{equation}
|F_l^{LP}(a)|\leqslant\Delta F(a).
\label{eqKM22}
\end{equation}
\noindent
Among these constraints, that ones on $\Lambda_2$ and $\Lambda_3$ turned out to be
stronger as compared with constraints found
from older Cavendish-type experiments available in 1987 \cite{KM73}.

It would be interesting to estimate potentialities of modern measurements of the Casimir
force for constraining the power-type interactions. For this purpose we consider the
most recent experiment \cite{KM33} on measuring the Casimir force between an Au-coted
sphere of $R=149.7~\mu$m radius and an Au-coated plate in the micrometer separation
range. The sphere is spaced at a height $a$ above the plate. To estimate the strongest
constraints that could be obtained from the experiments of this kind, we consider both
the sphere and the plate as all-gold (in real experiment the sapphire sphere and
silicon plate were coated with Au films of 250 and 150~nm thicknesses, respectively).
The plate can be considered as infinitely large because its size was much larger then
the sphere radius.

Let the plate top be in the plane $z=0$ and an atom of the sphere has the coordinates
${\mbox{\boldmath$r$}}_1=(0,0,z)$. For all powers $l\geqslant 3$ in (\ref{eqKM17}) the
plate can be considered as infinitely thick. The atom-plate force arising due to the
second contribution on the right-hand side of  (\ref{eqKM17}) is given by
\begin{eqnarray}
F_l^{AP}(a)&=&
Gm_1m_2n_2\Lambda_lr_0^{l-1}
\frac{\partial}{\partial z}
\int_{V_2}\!\!\!d^3 r_2 \frac{1}{|{\mbox{\boldmath$r$}}_1-{\mbox{\boldmath$r$}}_2|^{l-1}}
\nonumber \\
&=&-\frac{2\pi}{l-2}G\rho_2m_1\Lambda_lr_0^{l-1}\frac{1}{z^{l-2}},
\label{eqKM23}
\end{eqnarray}
\noindent
where $\rho_2=m_2n_2$ is the mass density of the plate material (Au).

Now we integrate (\ref{eqKM23}) over the volume of a sphere. The density of atoms at a height
$z\geqslant a$ in thin horizontal layer of the sphere is given by
\begin{equation}
\pi n_1\left[2R(z-a)-(z-a)^2\right].
\label{eqKM24}
\end{equation}

Then, the sphere-plate force is found by integrating (\ref{eqKM23}) with the weight
(\ref{eqKM24})
\begin{equation}
F_l^{SP}(a)=-\frac{2\pi^2}{l-2}G\rho_1\rho_2\Lambda_lr_0^{l-1}
\int_a^{2R+a}\frac{2R(z-a)-(z-a)^2}{z^{l-2}}\,dz,
\label{eqKM25}
\end{equation}
\noindent
where $\rho_1=m_1n_1$ is the mass density of the sphere material (in our case also Au).

Introducing the new integration variable $t=z-a$, we rewrite (\ref{eqKM25}) in the form
\begin{equation}
F_l^{SP}(a)=-\frac{2\pi^2}{l-2}G\rho_1\rho_2\Lambda_lr_0^{l-1}
\int_0^{2R}\frac{2Rt-t^2}{(a+t)^{l-2}}\,dt.
\label{eqKM26}
\end{equation}
\noindent
Finally, calculating the integral in (\ref{eqKM26}), one arrives at
\begin{equation}
F_l^{SP}(a)=-\frac{2\pi^2}{l-2}G\Lambda_l\rho_1\rho_2
\frac{r_0^{l-1}R^3}{a^{l-2}}{}_2F_1(l-2,2;3;-{2R}{a}^{-1}),
\label{eqKM27}
\end{equation}
\noindent
where ${}_2F_1(a,b;c;z)$ is the hypergeometric function.

Substituting (\ref{eqKM27}) in place of $F_l^{LP}$ in  (\ref{eqKM22}), one finds the
strongest constraints on $\Lambda_l$ obtainable from the experiment \cite{KM33}
if it would be performed with the all-gold test bodies. The numerical analysis shows
that the most strong constraints follow at $a=3~\mu$m where
$\Delta F(a)=2.2~$fN \cite{KM33}. The obtained constraints are:
$|\Lambda_3|\leqslant 1.3\times 10^{23}$,
$|\Lambda_4|\leqslant 1.8\times 10^{34}$, and
$|\Lambda_5|\leqslant 5.6\times 10^{44}$.
It is seen that these constraints are weaker than those mentioned above following
from the Cavendish-type experiment \cite{KM69,KM70}.

The case of $l=2$ should be considered separately. In this case, it is necessary
to take into account the finite thickness of the plate $D=50~\mu$m because for
$l=2$ an integral over the plate of infinitely large thickness (i.e., over the
semispace) diverges. By performing calculations in the same way as above, one obtains
\begin{eqnarray}
&&
F_2^{SP}(a)=-\frac{2\pi^2}{3}\rho_1\rho_2G\Lambda_2r_0\left[
\vphantom{\frac{a+2R+D}{a+2R}}
2RD(2a+2R+D)+(a+3R)a^2\ln\frac{a+2R}{a}
\right.
\nonumber \\
&&~~~~~~~~~~\left.
-(a+D)^2(a+3R+D)\ln\frac{a+2R+D}{a+D}+4R^3\ln\frac{a+2R+D}{a+2R}\right].
\label{eqKM28}
\end{eqnarray}
\noindent
Substituting  (\ref{eqKM28}) in (\ref{eqKM22}) in place of $F_l^{LP}$  and using
$\Delta F(a)=2.2~$fN, one finds
$|\Lambda_2|\leqslant 2.85\times 10^{12}$.
This is again a much weaker constraint than that obtained in \cite{KM70} based on
the Cavendish-type experiment \cite{KM69}. One can conclude that the short-separation
Cavendish-type experiments are more prospective for constraining the power-type
hypothetical interactions than measurements of the Casimir force.

This conclusion finds further confirmation from the recently performed
Cavendish-type experiment which presents an improved test of Newton's gravitational
law at short separations \cite{KM74}. The constraints on $\Lambda_l$ with $l=2$, 3, 4,
and 5 obtained in this work are somewhat stronger than those cited above \cite{KM69,KM70}.
In  Table~\ref{tb1} (line 1) we present the strongest constraint on $\Lambda_1$ following
from the E\"{o}tvos-type experiment \cite{KM67}. In columns 2 and 3 (lines 2--5), the
strongest constraints $\Lambda_l$ with $l=2$, 3, 4, and 5 following from the
Cavendish-type experiments \cite{KM70} and \cite{KM74}, respectively, are presented.
As is seen in Table~\ref{tb1}, the strength of constraints quickly drops with
increase of the interaction power.
\begin{table}
\caption{The strongest constraints on the constants of power-type hypothetical
interaction {\protect \\} following from the E\"{o}tvos-type (line 1) and Cavendish-type
(lines 2--5) experiments.}
\label{tb1}
\begin{tabular}{p{1.5cm}p{1.5cm}p{2cm}p{1.5cm}p{0.5cm}}
\hline\noalign{\smallskip}
~l & $|\Lambda_l|_{\max}$ & &$|\Lambda_l|_{\max}$&   \\
\noalign{\smallskip}\svhline\noalign{\smallskip}
~1		& $1\times 10^{-9}$	& \cite{KM67} & $1\times 10^{-9}$ & \cite{KM67}~\\
~2		& $4.5\times 10^{8}$ & \cite{KM70} & $3.7\times 10^{8}$ & \cite{KM74}~\\
~3		& $1.3\times 10^{20}$ & \cite{KM70} & $7.5\times 10^{19}$ & \cite{KM74}~\\
~4		& $4.9\times 10^{31}$ & \cite{KM70} & $2.2\times 10^{31}$ & \cite{KM74}~\\
~5		& $1.5\times 10^{43}$ & \cite{KM70} & $6.7\times 10^{42}$ & \cite{KM74}~\\
\noalign{\smallskip}\hline\noalign{\smallskip}
\end{tabular}
\vspace*{-12pt}
\end{table}

\section{Testing the Yukawa-Type Corrections to Newtonian Gravity from the Casimir Effect}
\label{secKM:4}

The interaction energy of Yukawa type between two pointlike particles (atoms or molecules)
separated by a distance $r$ arises due to an exchange of one light scalar particle.
The Standard Model considered in Sect.~\ref{secKM:1} contains only one scalar particle,
the Higgs boson, which is very heavy and cannot serve as an exchange boson in long-range
interactions. The extensions of the Standard Model predict, however, a number of light
scalar particles such as moduli \cite{KM75}, which arise in supersymmetric theories,
dilaton \cite{KM76}, which appears in extra-dimensional models with the varying volume
of compactified dimensions, scalar axion \cite{KM77}, which is a superpartner of an
axion, etc. (see Sect.~\ref{secKM:5} {and also the discussions on scalar-teleparallel and scalar-tensor theories in Chapters~\ref{Chap:0103_02TP}, \ref{Chap:0202_01_NS} and \ref{Chap:0202_04_Boson}).}

Similar to the power-type interactions, the interaction of Yukawa type between two
particles of masses $m_1$ and $m_2$ coexists with gravity and is usually parametrized
as
\begin{equation}
V_{\rm Yu}(r)=-\frac{Gm_1m_2}{r}\left(1+\alpha e^{-r/\lambda}\right),
\label{eqKM29}
\end{equation}
\noindent
where $\alpha$ is the dimensionless interaction constant and $\lambda$ is the interaction
range having the meaning of the Compton wavelength of exchange scalar particle of mass
$m_s$: $\lambda=\hbar/(m_sc)$.

Another prediction of the Yukawa-type correction to Newton's gravitational law comes from
the extra-dimensional models with compact extra dimensions and low-energy compactification
scale \cite{KM78,KM79}. In the framework of this approach beyond the Standard Model,
the space-time has $D=4+N$ dimensions where $N$ extra dimensions are compactified  at
relatively low Planck energy scale in $D$ dimensions
\begin{equation}
E_{\rm Pl}^{(D)}=\left(\frac{\hbar^{1+N}c^{5+N}}{G_D}\right)^{\frac{1}{2+N}}
\sim 1~\mbox{TeV}.
\label{eqKM30}
\end{equation}
\noindent
Here, $G_D$ is the gravitational constant in the extended $D$-dimensional space-time
$G_D=G\Omega_N$ and $\Omega_N\sim R_{\ast}^N$, $R_{\ast}$ being the size of compact manifold.

In fact the approach under consideration was suggested as a possible solution of
hierarchy problem discussed in Sect.~\ref{secKM:1} since due to (\ref{eqKM30}) the
characteristic energy scales of the gravitational and gauge interactions of the Standard
Model coincide. In doing so the size of compact manifold is given by \cite{KM79}
\begin{equation}
R_{\ast}\sim\frac{\hbar c}{E_{\rm Pl}^{(D)}}\left[
\frac{E_{\rm Pl}}{E_{\rm Pl}^{(D)}}\right]^{\frac{2}{N}}\sim
10^{\frac{32-17N}{N}},
\label{eqKM31}
\end{equation}
\noindent
where the usual Planck energy $E_{\rm Pl}=(\hbar c^5/G)^{1/2}\sim 10^{19}~$GeV.

According to the developed approach, the standard Newton law (\ref{eqKM1}) and  (\ref{eqKM2})
is not valid in $D$-dimensional space-time. It was shown \cite{KM80,KM81} that at separations
$r\gg R_{\ast}$ the gravitational interaction energy takes the form (\ref{eqKM29}) with
$\lambda\sim R_{\ast}$. Although for one extra dimension ($N=1$) equation (\ref{eqKM31})
leads to too large $R_{\ast}\sim 10^{15}~$cm, which is excluded by the tests of Newton's
law in the solar system \cite{KM82}, for $N=2$ and 3 (\ref{eqKM31}) leads to the more
realistic results $R_{\ast}\sim 1~$mm and $R_{\ast}\sim 5~$nm, respectively.
This means that a search for deviations from the Newton law at short distances is not only
a quest for hypothetical particles, but  for extra dimensions as well.

Constraints on the parameters of Yukawa-type interaction $\alpha$ and $\lambda$ can be
obtained from the Cavendish-type experiments. The potential energy (\ref{eqKM29}) results in
the force acting between two particles $m_1$ and $m_2$
\begin{equation}
F_{\rm Yu}(r)=-\frac{d V_{\rm Yu}(r)}{d r}=
-\frac{Gm_1m_2}{r^2}\left[1+\alpha e^{-r/\lambda}\left(
1+\frac{r}{\lambda}\right)\right].
\label{eqKM32}
\end{equation}
\noindent
Then the  quantity
\begin{equation}
\varepsilon_{\rm Yu}=\frac{1}{rF_{\rm Yu}(r)}
\frac{d}{d r}\left[r^2F_{\rm Yu}(r)\right]
\label{eqKM33}
\end{equation}
\noindent
is not equal to zero due to a nonzero strength of the Yukawa force $\alpha$.
The deviation of this quantity from zero (if any)
could be determined from the results of Cavendish-type experiments.
Depending on the range of $\lambda$, different Cavendish-type experiments lead to
the strongest constraints on $\alpha$.
For $8~\mu\mbox{m}<\lambda<9~\mu$m the most strong constraints follow from the
short-range test of Newtonian gravity at 20 micrometers \cite{KM83}.
The Cavendish-type experiment \cite{KM69}, already discussed in Sect.~\ref{secKM:3}
in the context of power-type interactions, leads to the strongest constraints on $\alpha$
within the wide interaction range $9~\mu\mbox{m}<\lambda<4~$mm \cite{KM70}.
It should be noted, however, that in the part of this interval
$40~\mu\mbox{m}<\lambda<0.35~$mm the obtained results have been strengthened by up to
a factor of 3 in the refined experiment \cite{KM74} which was also mentioned in
Sect.~\ref{secKM:3}. Finally, an older Cavendish-type experiment \cite{KM84} performed
at larger separations allows obtaining the strongest constraints on $\alpha$ in the
range $4~\mbox{mm}<\lambda<1~$cm. In the range of even larger $\lambda$, unrelated to
the Casimir force, the strongest constraints on $\alpha$  follow from the E\"{o}tvos-type
experiments \cite{KM68,KM85}.

\begin{figure}[b]
\sidecaption
\includegraphics[scale=.35]{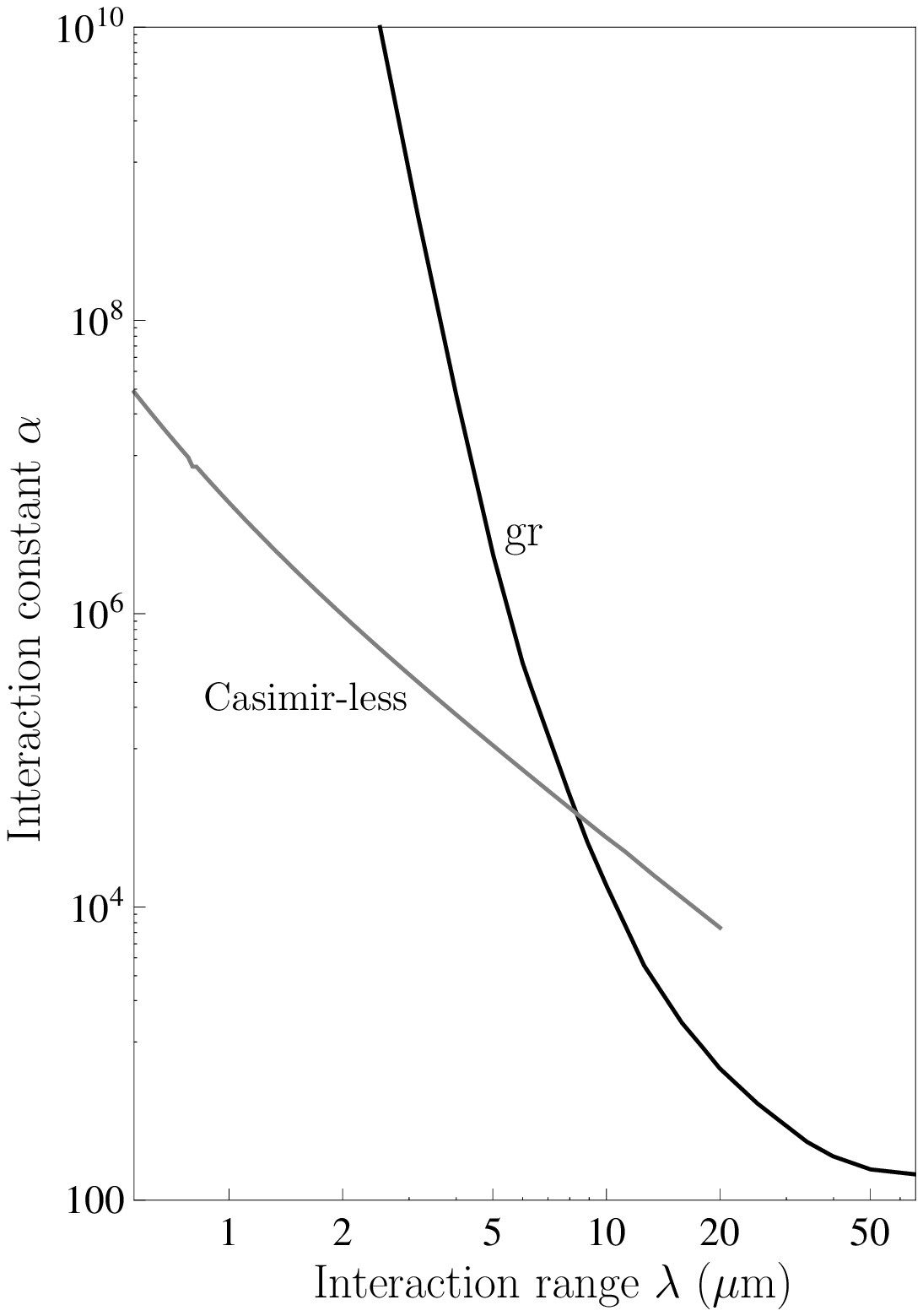}
\caption{Constraints on the interaction constant $\alpha$ of Yukawa-type interaction
are shown as functions of the interaction range $\lambda$ by the lines labeled gr and
Casimir-less obtained from the gravitational and Casimir-less experiments, respectively.
The regions of $(\lambda,\alpha)$ plane above each line are excluded and below are
allowed}
\label{figKM:1}
\end{figure}
In Fig.~\ref{figKM:1}, we present the constraints on $\alpha$ obtained from different
 gravitational experiments by the line labeled gr. Only the range of $\lambda$ below
$66~\mu$m is included neighboring to the region considered below where the strongest
constraints on $\alpha$ follow from experiments performed in the Casimir regime.
The values of parameters of the Yukawa-type interaction belonging to the area of
$(\lambda,\alpha)$ plane above the line are excluded by the results of Cavendish-type
experiments mentioned above, whereas the area of the same plane below the line is
allowed. At $\lambda=8~\mu$m the line gr intersects with the end of the line
Casimir-less which is discussed below in this section.

As is seen in Fig.~\ref{figKM:1}, the strength of constraints obtained from gravitational
experiments quickly drops with decreasing $\lambda$. As an example, for $r=\lambda=10~\mu$m
the Cavendish-type experiments do not exclude an existence of the Yukawa-type force
between two particles which exceeds the Newtonian gravitational force by the factor of
$10^4$. This means that the Newtonian law of gravitation lacks of sufficient experimental
confirmation at short separations which prevents obtaining strong constraints on some
other forces from gravitational experiments.

In fact at separations of the order of micrometer and less the main background force
between two material bodies far exceeding the gravitational interaction is the Casimir
force considered in Sect.~\ref{secKM:2}. In \cite{KM86} it was suggested to constrain
the hypothetical Yukawa-type interaction from experiments on measuring the van der Waals
and Casimir forces.

Similar to the case of power-type interactions, the Yukawa-type force acting between two
test bodies spaced at a closest separation $a$ can be obtained by an integration of
the interaction energy (\ref{eqKM29}) over their volumes with subsequent negative
differentiation with respect to $a$
\begin{equation}
F_{\rm Yu}^{V_1V_2}(a)=-n_1n_2\frac{\partial}{\partial a}
\int_{V_1}\!\!\!d^3r_1\int_{V_2}\!\!\!d^3r_2
V_{\rm Yu}(|{\mbox{\boldmath$r$}}_1-{\mbox{\boldmath$r$}}_2|).
\label{eqKM34}
\end{equation}

Taking into account that within the experimental error $\Delta F(a)$ the measured
Casimir force was found to be in agreement with theoretical predictions, the
constraints on $F_{\rm Yu}$ can be found from the inequality
\begin{equation}
|F_{\rm Yu}^{V_1V_2}(a)|\leqslant\Delta F(a).
\label{eqKM35}
\end{equation}
\noindent
Following this approach, the first constraints on the Yukawa-type interaction with
$\lambda<20~$cm were obtained \cite{KM86} from two experiments \cite{KM72,KM87}
performed long ago.

During the last twenty years many experiments on measuring the Casimir interaction
have been performed (some of them are mentioned in Sect.~\ref{secKM:2}). All these
experiments use the configuration of a sphere above a plate which surfaces may be
coated by some additional material layers or covered with sinusoidal corrugations.
We start with the simplest configuration of a smooth sphere of radius $R$ at the
closest separation $a$ above a large smooth plate of thickness $D$.
We again consider an atom $m_1$ of the sphere at a height $z$ above the plate and
integrate the Yukawa interaction energy (\ref{eqKM29}) over the plate volume $V_2$
with subsequent negative differentiation according to (\ref{eqKM34}).
As explained above, the contribution of gravitational interaction can be neglected.
Then, similar to (\ref{eqKM23}), for the atom-plate force one obtains
\begin{equation}
F_{\rm Yu}^{AP}(z)=-2\pi G\rho_2m_1\alpha\lambda e^{-z/\lambda}\left(1-
e^{-D/\lambda}\right).
\label{eqKM36}
\end{equation}

Now we integrate (\ref{eqKM36}) over the sphere volume using (\ref{eqKM24}) and
obtain the Yukawa-type force acting between a sphere and a plate \cite{KM88}
\begin{eqnarray}
F_{\rm Yu}^{AP}(a)&=&-2\pi^2 G\rho_1\rho_2\alpha\lambda \left(1-
e^{-D/\lambda}\right)
\int_a^{2R+a}\!\!\!dz\left[2R(z-a)-(z-a)^2\right] e^{-z/\lambda}
\nonumber \\
&=&-4\pi^2 G\rho_1\rho_2\alpha\lambda^3 \left(1-e^{-D/\lambda}\right)
e^{-a/\lambda} \Phi(R.\lambda),
\label{eqKM37}
\end{eqnarray}
\noindent
where the following notation is introduced
\begin{equation}
\Phi(r,\lambda)=r-\lambda+(r+\lambda)e^{-2r/\lambda}.
\label{eqKM38}
\end{equation}

The strongest current constraints on the Yukawa-type force in the wide interaction range from
10~nm to $8~\mu$m follow from four experiments of the Casimir physics.

The first of them is devoted to measurements of the lateral Casimir force between the surfaces of a sphere and
a plate covered with coaxial longitudinal sinusoidal corrugations and coated with an Au film
\cite{KM89,KM90}. This experiment was performed by means of an atomic force microscope.

\begin{figure}[t]
\sidecaption
\includegraphics[scale=.35]{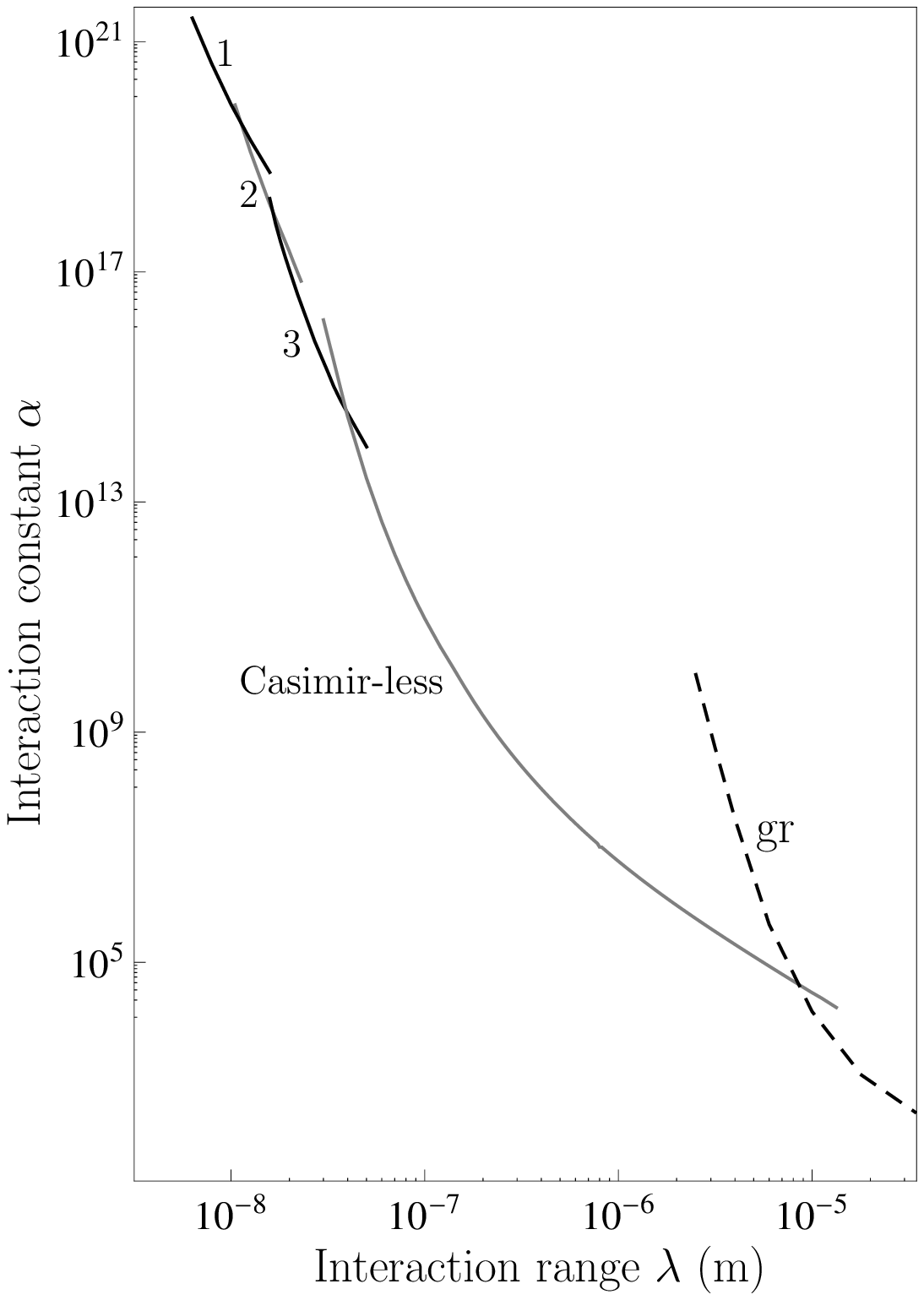}
\caption{Constraints on the interaction constant $\alpha$ of Yukawa-type interaction
are shown as functions of the interaction range $\lambda$ by the lines labeled 1, 2, 3,
Casimir-less, and gr  obtained from measuring the lateral and normal Casimir forces
between the sinusoidally corrugated surfaces, effective Casimir pressure, from the
Casimir-less experiment, and gravitational experiments, respectively.
The regions of $(\lambda,\alpha)$ plane above each line are excluded and below are
allowed}
\label{figKM:2}
\end{figure}

The respective constraints were obtained in \cite{KM91}. For this purpose, the interaction
energy of Yukawa type between the corrugated test bodies was calculated by integrating
(\ref{eqKM29}) over their volumes, and the lateral force was found by the negative
differentiation of the obtained result with respect to the phase shift between corrugations
(see \cite{KM91} for details). The obtained constraints cover a wide interaction range
but currently they are the strongest ones only in the narrow region
$10~\mbox{nm}<\lambda<11.6~$nm (see the line labeled 1 in Fig.~\ref{figKM:2}).

The second experiment, also performed by using an atomic force microscope, is on measuring
the usual (normal) Casimir force between the sinusoidally corrugated Au-coated surfaces of
a sphere and a plate under some angle between the corrugation axes \cite{KM92,KM93}.
The constraints on the Yukawa parameters $\alpha$ and $\lambda$, following from this
experiment, were obtained in \cite{KM94}. Currently, these constraints remain the strongest
ones in the region $11.6~\mbox{nm}<\lambda<17.2~$nm.
They are shown by the line labeled 2 in Fig.~\ref{figKM:2}.

In the next, third, experiment the effective Casimir force per unit area of two Au-coated
plates (i.e., the effective Casimir pressure) was determined by means of a micromechanical
torsional oscillator \cite{KM27,KM28}. In fact it was recalculated from the directly
measured gradient of the Casimir force, $F_{sp}^{\prime}$, between a sphere and a plate
using (\ref{eqKM14}). In the same way, calculating the gradient of (\ref{eqKM37}) one
finds from (\ref{eqKM14}) the Yukawa-type pressure between two parallel plates
\begin{equation}
P_{\rm Yu}(a)=-2\pi G\rho_1\rho_2\alpha\lambda^2 \left(1-e^{-D/\lambda}\right)
e^{-a/\lambda}.
\label{eqKM39}
\end{equation}
\noindent
Here, following \cite{KM27,KM28}, we took into account that $\lambda\ll R$ leading
to $\Phi(R,\lambda)\approx R$.

In this experiment, the test bodies were not homogeneous. A sapphire sphere of density
$\rho_s=4.1~\mbox{g/cm}^3$ was coated with the first layer of Cr with density
$\rho_{\rm Cr}=7.14~\mbox{g/cm}^3$ and thickness $\Delta_1=10~$nm, and then with the
second, external, layer of Au of  density
$\rho_{\rm Au}=19.28~\mbox{g/cm}^3$ and thickness $\Delta_2=180~$nm.
The plate was made of Si with density $\rho_{\rm Si}=2.33~\mbox{g/cm}^3$ and coated
with a layer of Cr of thickness $\Delta_1=10~$nm and external layer of Au of
thickness $\widetilde{\Delta}_2=210~$nm.
Taking into account that the layers contribute to the Casimir pressure additively, we
obtain from (\ref{eqKM39}) the following expression valid in the experimental
configuration
\begin{eqnarray}
P_{\rm Yu}(a)&=&-2\pi G\alpha\lambda^2 e^{-a/\lambda}
\left[\rho_{\rm Au}-(\rho_{\rm Au}-\rho_{\rm Cr})e^{-\Delta_2/\lambda}-
(\rho_{\rm Cr}-\rho_{s})e^{-(\Delta_2+\Delta_1)/\lambda}\right]
\nonumber \\
&\times&
\left[\rho_{\rm Au}-(\rho_{\rm Au}-\rho_{\rm Cr})e^{-\widetilde{\Delta}_2/\lambda}-
(\rho_{\rm Cr}-\rho_{\rm Si})e^{-(\widetilde{\Delta}_2+\Delta_1)/\lambda}\right].
\label{eqKM40}
\end{eqnarray}

The constraints on the Yukawa parameters $\alpha$ and $\lambda$ were obtained from the
inequality
\begin{equation}
|P_{\rm Yu}(a)|\leqslant\Delta P(a),
\label{eqKM41}
\end{equation}
\noindent
where the experimental error $\Delta P(a)$ in measuring the effective Casimir pressure,
with which the theoretical predictions of the Lifshitz theory were confirmed, was
determined at the 95\% confidence level. Currently, the constraints obtained from this
experiment are the strongest ones over the interaction range
$17.2~\mbox{nm}<\lambda<39~$nm. They are found from (\ref{eqKM41}) at $a=180~$nm where
$\Delta P(a)=4.8~$mPa \cite{KM27,KM28} and
shown by the line labeled 3 in Fig.~\ref{figKM:2}.

The last, fourth, experiment leading to the strongest constraints on the Yukawa-type
interaction over a wide interaction range $39~\mbox{nm}<\lambda<8~\mu$m is performed
in such a way, that the contribution of the Casimir force to the measured signal is
nullified \cite{KM95}. This was achieved by measuring the differential force between
a sphere of $R=149.3~\mu$m radius and an especially structured plate using
a micromechanical torsional oscillator.
The sphere made of sapphire was coated with a $\Delta_1=10~$nm layer of Cr and
$\Delta_2=250~$nm layer of Au. The plate consisted of Si and Au parts of $D=2.1~\mu$m
thickness both coated with a Cr and Au overlayers of thicknesses $\Delta_1=10~$nm and
$\widetilde{\Delta}_2=150~$nm, respectively.

The Casimir forces between a sphere and two halves of the patterned Au-Si
plate are equal because the thickness of an Au overlayer is sufficiently large in order
it could be considered as a semispace \cite{KM14}.  As a result, when the sphere is
moved back and forth above the patterned plate, the measured differential force is equal
to a difference of the Yukawa-type forces between a sphere and two halves of the plate.
Using (\ref{eqKM37}) with $\Phi(R,\lambda)\approx R$ and taking into account the Au and Cr
layers covering the test bodies, the differential Yukawa-type force takes the form
\begin{eqnarray}
&&
F_{\rm Yu,Au}^{SP}(a)-F_{\rm Yu,Si}^{SP}(a)=-4\pi^2 G\alpha\lambda^3R
(\rho_{\rm Au}-\rho_{\rm Si}) e^{-(a+\widetilde{\Delta}_2+\Delta_1)/\lambda}
\left(1-e^{-D/\lambda}\right)
\nonumber \\
&&~~~~~~~~~~~~\times
\left[\rho_{\rm Au}-(\rho_{\rm Au}-\rho_{\rm Cr})e^{-\Delta_2/\lambda}-
(\rho_{\rm Cr}-\rho_{s})e^{-(\Delta_2+\Delta_1)/\lambda}\right].
\label{eqKM42}
\end{eqnarray}

The constraints have been obtained from the
inequality
\begin{equation}
|F_{\rm Yu,Au}^{SP}(a)-F_{\rm Yu,Si}^{SP}(a)|\leqslant\Xi(a),
\label{eqKM43}
\end{equation}
\noindent
where  a sensitivity of the setup to force differences $\Xi(a)$ is equal to a fraction
of 1~fN. Note that both the residual electric and Newtonian gravitational forces contribute
well below this sensitivity \cite{KM95}. The strongest current constraints of $\alpha$ and
$\lambda$ obtained from (\ref{eqKM43}) extend over a wide interaction range
$40~\mbox{nm}<\lambda<8~\mu$m (see the line labeled Casimir-less in Fig.~\ref{figKM:2}).

Thus, Fig.~\ref{figKM:2} presents the strongest constraints on the Yukawa-type interaction
obtained from Casimir physics. Almost all these constraints with except of the region
from 10 to 39~nm are obtained in \cite{KM95} which is the fourth experiment discussed above.
At $\lambda=8~\mu$m the constraints found from the differential force measurements
(which are also called the Casimir-less experiment) are of the same strength as the
constraints found from the Cavendish-type experiments. At larger $\lambda$ the strongest
constraints are shown by a beginning of the line labeled gr reproduced from
Fig.~\ref{figKM:1}.

\begin{figure}[t]
\sidecaption
\includegraphics[scale=.35]{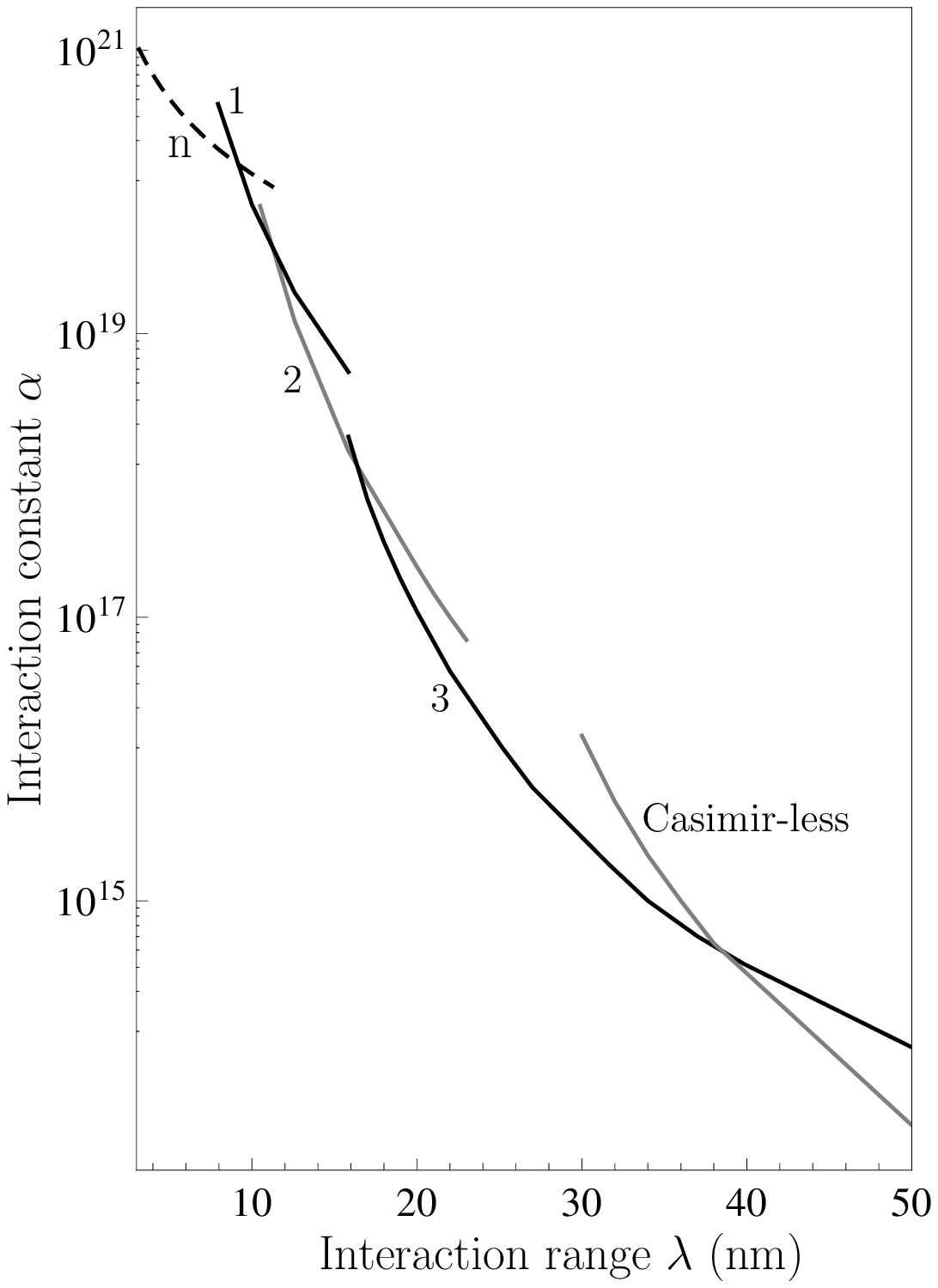}
\caption{Constraints on the interaction constant $\alpha$ of Yukawa-type interaction
are shown as functions of the interaction range $\lambda$ by the lines labeled n, 1, 2, 3,
and Casimir-less,   obtained from the experiments on neutron scattering,
measuring the lateral and normal Casimir force
between the sinusoidally corrugated surfaces, effective Casimir pressure, and from the
Casimir-less experiment,  respectively.
The regions of $(\lambda,\alpha)$ plane above each line are excluded and below are
allowed}
\label{figKM:3}
\end{figure}
In Fig.~\ref{figKM:3} we reproduce the beginning of the line labeled Casimir-less in
Fig.~\ref{figKM:2} on an enlarged scale in order to better demonstrate the constraints
obtained in the range $10~\mbox{nm}<\lambda<39~$nm \cite{KM27,KM28,KM91,KM94} from
the experiments \cite{KM27,KM28,KM89,KM90,KM92,KM93} (i.e., from the first, second  and
third experiments discussed above). In this figure, the lines labeled n also indicate the
strongest constraints on the Yukawa-type interaction obtained at $\lambda<10~$nm from
the experiments on neutron scattering \cite{KM96,KM97} (in the region
$0.03~\mbox{nm}<\lambda<0.1~$nm the strongest constraints follow from the experiment
using a pulsed neutron beam \cite{KM98}).

There are also many other papers in the scientific literature devoted to constraining
the Yukawa-type corrections to Newtonian gravity from the Casimir effect (see, e.g.,
\cite{KM99,KM100,KM101,KM102,KM103}). The constraints obtained there are, however,
somewhat weaker than the current strongest constraints presented in  Figs.~\ref{figKM:2}
and \ref{figKM:3}. It should be mentioned that in the range of extremely small
$\lambda$ the constraints on $\alpha$ have been obtained from spectroscopic measurements
in simple atomic systems like hydrogen and deuterium whose spectra can be calculated
and measured with high precision. Thus, it was found that in the range
$2\times 10^{-4}~\mbox{nm}<\lambda<20~$nm the maximum strength of Yukawa interaction
varies from $2\times 10^{27}$ to  $2\times 10^{25}$ \cite{KM104}.

\section{Dark Matter, Dark Energy and Their Hypothetical Constituents}
\label{secKM:5}

According to astrophysical observations, the visible matter in the
form of stars, galaxies, planets and radiation constitutes only
about 5\% of the total mass of the Universe. By studying stellar
motion in the neighborhood of our galaxy ninety years ago, Oort
found \cite{KM105} that the galaxy mass must be much larger than
the mass of all stars belonging to it. At the same time, an
application of the virial theorem to the Coma cluster of galaxies
by Zwicky \cite{KM106} resulted in a much larger mass than that
found by summing up the masses of all observed galaxies belonging
to this cluster. In succeeding years, these results received ample
recognition. Presently it is generally agreed that the dark matter,
which reveals itself only gravitationally, is not composed of
elementary particles of the Standard Model listed in Sect.~\ref{secKM:1}
and adds up approximately 27\% of the Universe energy.

The problem of what dark matter is remains unresolved. There are
many approaches to its resolution which consider some hypothetical
particles introduced in different theoretical schemes beyond the
Standard Model as possible constituents of dark matter. Among these
particles are axions, arions, massive neutrinos, weakly interacting
massive particles (WIMP) etc. The possibility to explain the
observational data by a modification of the gravitational theory in
place of compensating for a deficiency in matter is also
investigated. All these approaches are widely discussed in the
literature \cite{KM107,KM108,KM109,KM110,KM111}.

During the last few years, the major support from astrophysics and
cosmology was received by the model of cold dark matter. This model
suggests that the constituents of dark matter are light particles
which were produced at the first stages of the Universe evolution
and became nonrelativistic long ago. The best candidate of this kind
is a pseudoscalar {Nambu}-Goldstone boson called an axion.

This particle was introduced \cite{KM77,KM112,KM113} for solving the
problem of strong CP violation in quantum chromodynamics mentioned
in Sect.~\ref{secKM:1}, i.e., independently of the problem of dark matter. The
point is that all the experimental data show that strong interactions
are CP invariant and the electric dipole moment of a neutron is equal
to zero. In contrast to these facts, the vacuum state of quantum
chromodynamics depends on an angle $\theta$ which violates the CP
invariance and allows a nonzero electric dipole moment of a neutron.
To resolve this contradiction between experiment and theory,
Peccei and Quinn \cite{KM77} introduced the new symmetry which
received their names. In doing so the emergence of axions is a direct
consequence of the violation of this symmetry \cite{KM112,KM113}.

Later it was understood that axions and other axionlike particles
arise in many extensions of the Standard Model. They can interact
with particles of the Standard Model, e.g., with photons, electrons
and nucleons, and lead to a number of processes which could be
observed both in the laboratory experiments and in astrophysics and
cosmology (see
\cite{KM107,KM108,KM109,KM110,KM111,KM114,KM115,KM116,KM117,KM118,KM119,KM120}
for a review). The question arises whether it is possible to constrain
the parameters of axions from the Casimir effect. This question is
considered below in Sects.~\ref{secKM:6} and \ref{secKM:7}.

Another unresolved problem of modern physics is the problem of dark
energy. In the end of twentieth century, the observations of
supernovae demonstrated that an expansion of the Universe is
accelerating \cite{KM121}. This fact is in some contradiction with
expectations based on the general relativity theory and the properties
of matter described by the Standard Model because the gravitational
interaction of usual matter is attractive and should make the Universe
expansion slower.

The concept of dark energy, i.e., some new kind of invisible matter
which causes a repulsion, was introduced in numerous discussions of
this problem. One approach to its resolution goes back to Einstein's
cosmological constant which is closely connected with the problem of
the quantum vacuum. According to the observational data, the dark
energy constitutes as much as approximately 68\% of the Universe
ener\-gy. This corresponds to some background medium (physical
vacuum) posses\-sing the energy density of $\epsilon \approx 10^{-9}$
J/m$^3$. In order that this medium could accelerate the Universe
expansion, it should possess the equation of state $P=-\epsilon$<0,
i.e., the negative pressure.

The cosmological term $\Lambda g_{ik}$, where $g_{ik}$ is the metrical
tensor, when added to Einstein's equations of general relativity
theory, results in just this equation of state. In doing so, the value
of $\Lambda$ is determined by the above value of $\epsilon$ determined
from the observed acceleration of the Universe expansion
\begin{equation}
\Lambda=8\pi G\epsilon\approx 2\times 10^{-52} \mbox{m}^{-2}.
\label{eqKM44}
\end{equation}

It was argued, however, that quantum field theory using a cutoff at
the Planck momentum $p_{Pl}=E_{Pl}/c$ leads to quite a different
value of the vacuum energy density $\epsilon_{vac}\approx 10^{111}$J/m$^3$
which is different from $\epsilon$ determined from observations by the
factor of $10^{120}$ \cite{KM122,KM123}. If to take into account that
the vacuum energy density admits an interpretation in term of the
cosmological constant \cite{KM124}, it becomes clear why this
discrepancy by the factor of $10^{120}$ was called the vacuum
catastrophe \cite{KM125}.

Another approach to the understanding of dark energy attempts to model it
by the fields and respective particles with unusual physical properties.
One of the models of this kind introduces the real self-interacting
scalar field $\phi$ with a variable mass called chameleon \cite{KM126}.
The distinctive feature of chameleon particles is that they become
heavier in more dense environments and lighter in free space.

Another model similar in spirit suggests that the interaction constant of
the self-interacting real scalar field with usual matter depends on the
density in the environment. The fields and particles of this kind are
called symmetrons \cite{KM127,KM128,KM129}. Symmetrons interact with
usual matter described by the Standard Model weaker if the density of the
environment is higher.

There are also other hypothetical particles which could lead to the
negative pressure and help to understand the accelerating expansion
of the Universe. For instance, the negative pressure originates from the
Maxwell stress-energy tensor of massive photons in the Maxwell-Proca
electrodynamics \cite{KM130}.

If the exotic particles, such as chameleons, symmetrons, massive
photons etc., exist in nature, this should lead to some additional
forces between the closely spaced macrobodies. In Sect.~\ref{secKM:8}
the possibility of constraining these forces from measurements of the
Casimir force is discussed.

\section{Constraining Dark Matter Particles from the Casimir Effect}
\label{secKM:6}

As was mentioned in previous section, the main candidate for the role of a dark matter particle
is light pseudoscalar particle called an axion which can interact with photons, electrons,
and nucleons. It can be easily seen that the interaction of axions with photons and
electrons does not lead to sufficiently large forces between the closely spaced bodies
which could be constrained from measurements of the Casimir force. These interactions of
axions are investigated by other means. For example, the conversion process of photons
into axions in strong magnetic field (the so-called Primakoff process) is used for an
axion search in astrophysics \cite{KM131} (see also reviews \cite{KM116,KM117,KM118,KM119}
for already obtained constraints on interactions of axions with photons and electrons).

Here, we concentrate our attention on the interaction of axions with nucleons (neutrons and protons) which could lead to some noticeable additional force between two neighboring
bodies. The interaction Lagrangian density between the originally introduced axion field
$a(x)$ and the fermionic field $\psi(x)$ is given by \cite{KM114,KM117}
\begin{equation}
{\cal L}_{pv}(x)=\frac{g}{2m_a}\hbar^2\bar{\psi}(x)\gamma_5\gamma_{\mu}\psi a(x)
\partial^{\mu}a(x),
\label{eqKM45}
\end{equation}
\noindent
where $g$ is the dimensionless interaction constant, $m_a$ is the axion mass, $\gamma_{\mu}$
with $\mu=0$, 1, 2, 3 and $\gamma_5$ are the Dirac matrices.
The Lagrangian density (\ref{eqKM45}) is called pseudovector. It describes the interaction
of fermions with pseudo Nambu-Goldstone bosons.

Various extensions of the Standard Model called the Grand Unified Theories (GUT) introdice
the axionlike particles which interact with fermions through the pseudoscalar
Lagrangian density  \cite{KM114,KM117,KM132}
\begin{equation}
{\cal L}_{ps}(x)=-i{g}\hbar c\bar{\psi}(x)\gamma_5\psi(x)a(x).
\label{eqKM46}
\end{equation}
\noindent
Unlike (\ref{eqKM45}), which contains a dimensional effective interaction constant $g/m_a$,
the Lagrangian density (\ref{eqKM46}) results in a renormalizable field theory.

When one considers an exchange of a single axion between two nucleons of mass $m$ belonging
to the closely spaced test bodies, both Lagrangian densities (\ref{eqKM45}) and
(\ref{eqKM46}) lead to the common effective potential energy \cite{KM133,KM134}
\begin{eqnarray}
V_{an}(r;\mbox{\boldmath$\sigma$}_1,\mbox{\boldmath$\sigma$}_2)&=&
\frac{g^2\hbar^3}{16\pi m^2c}\left[(\mbox{\boldmath$\sigma$}_1\cdot\mbox{\boldmath$n$})
(\mbox{\boldmath$\sigma$}_2\cdot\mbox{\boldmath$n$})
\left(\frac{m_a^2c^2}{\hbar^2r}+\frac{3m_ac}{\hbar r^2}+\frac{3}{r^3}\right)\right.
\nonumber \\
&-&\left.
(\mbox{\boldmath$\sigma$}_1\cdot\mbox{\boldmath$\sigma$}_2)
\left(\frac{m_ac}{\hbar r^2}+\frac{1}{r^3}\right)\right],
\label{eqKM47}
\end{eqnarray}
\noindent
where $r=|\mbox{\boldmath$r$}_1-\mbox{\boldmath$r$}_2|$ is a distance between nucleons,
$\mbox{\boldmath$\sigma$}_1,{\ }\mbox{\boldmath$\sigma$}_2$ are their spins, and
$\mbox{\boldmath$n$}=(\mbox{\boldmath$r$}_1-\mbox{\boldmath$r$}_2)/r$
is the unit vector along the line connecting these nucleons.

The effective interaction energy (\ref{eqKM47}) depends on the spins of nucleons and the
respective force averages to zero after a summation over the volumes of unpolarized test
bodies. Because of this, using (\ref{eqKM47}), the parameters of axion $g$ and $m_a$ can
not be constrained from experiments on measuring the Casimir force discussed in
Sect.~\ref{secKM:2} (the possibilities of constraining the spin-dependent interactions
are considered in the next section).

There is, however, the possibility to obtain the spin-independent interaction energy
between two nucleons by considering the process of two-axion exchange.
If the Lagrangian density (\ref{eqKM46}) is used, the effective interaction energy is
given by \cite{KM82,KM135,KM136}
\begin{equation}
V_{aan}(r)=-\frac{g^4\hbar^2}{32\pi^3m^2}\,\frac{m_a}{r^2}
K_1\left(\frac{2m_acr}{\hbar}\right),
\label{eqKM48}
\end{equation}
\noindent
where $K_1(z)$ is the modified Bessel function of the second kind.

In the case of Lagrangian density (\ref{eqKM45}), the respective field theory is
nonrenormalizable. As a result, the effective interaction energy between nucleons
due to an exchange of two axions is not yet available (see \cite{KM137} for more details).
This means that measurements of the Casimir force can be used for constraining only the
parameters of GUT axions described by the pseudoscalar Lagrangian density (\ref{eqKM46}).

Similar to the cases of power-type and Yukawa-type interactions in (\ref{eqKM21}) and
(\ref{eqKM34}), the hypothetical force between two experimental test bodies due to
two-axion exchange of their nucleons is given by
\begin{equation}
F_{aan}(a)=-n_1n_2\frac{\partial}{\partial a}
\int_{V_1}\!\!\!d^3r_1\int_{V_2}\!\!\!d^3r_2
V_{aan}(|{\mbox{\boldmath$r$}}_1-{\mbox{\boldmath$r$}}_2|),
\label{eqKM49}
\end{equation}
\noindent
where $a$ is the closest distance between these bodies and $n_1,\,n_2$ are the numbers
of nucleons per unit volume of their materials.

We consider first a homogeneous Au sphere above a homogeneous Si plate of thickness $D$
which is assumed to be infinitely large. Substituting (\ref{eqKM48}) in (\ref{eqKM49}) and
using the integral representation \cite{KM138}
\begin{equation}
\frac{K_1(z)}{z}=\int_1^{\infty}\!\!\!du\sqrt{u^2-1}e^{-zu},
\label{eqKM50}
\end{equation}
\noindent
one obtains
\begin{eqnarray}
F_{aan}^{SP}(a)&=&-\frac{\pi m_a\hbar^2}{m^2m_{\rm H}^2}C_1C_2
\int_1^{\infty}\!\!\!du\frac{\sqrt{u^2-1}}{u}\left(1-e^{-2m_acuD/\hbar}\right)
\nonumber \\
&\times&
\int_a^{2R+a}\!\left[2R(z-a)-(z-a)^2\right]e^{-2m_acuz/\hbar}dz.
\label{eqKM51}
\end{eqnarray}
\noindent
Here, the coefficients $C_1$ and $C_2$ are defined for a sphere and a plate materials,
respectively, in the following way:
\begin{equation}
C_{1,2}=\rho_{1,2}\frac{g_{an}^2}{4\pi}\left(\frac{Z_{1,2}}{\mu_{1,2}}+
\frac{N_{1,2}}{\mu_{1,2}}\right),
\label{eqKM52}
\end{equation}
\noindent
where  $Z_{1,2}$ and $N_{1,2}$ are the numbers of protons and the mean number of neutrons
in the sphere and plate atoms, respectively, and $\mu_{1,2}=m_{1,2}/m_{\rm H}$ are defined
as the mean masses of a sphere and a plate atoms divided by the mass of atomic hydrogen.

By integrating in (\ref{eqKM51}) with respect to $z$, one obtains
\begin{eqnarray}
F_{aan}^{SP}(a)&=&-\frac{\pi\hbar^4}{2m_am^2m_{\rm H}^2c^2}C_1C_2
\int_1^{\infty}\!\!\!du\frac{\sqrt{u^2-1}}{u^3}e^{-2m_acua/\hbar}
\nonumber \\
&\times&
\left(1-e^{-2m_acuD/\hbar}\right)
\,\chi\!\left(R,\frac{m_acu}{\hbar}\right),
\label{eqKM53}
\end{eqnarray}
\noindent
where the function $\chi(r,z)$ similar to $\Phi(r,\lambda)$ in (\ref{eqKM38}) is defined as
\begin{equation}
\chi(r,z)=r-\frac{1}{2z}+\left(r+\frac{1}{2z}\right)e^{-4rz}.
\label{eqKM54}
\end{equation}

The strongest constraints on the parameters of axionlike particles were obtained \cite{KM139}
from the differential measurements where the contribution of the Casimir force was
nullified \cite{KM95}. This experiment was already discussed in Sect.~\ref{secKM:4}.
Taking into account the structure of the plate consisting of Au and Si halves, as well as
additional Cr and Au layers (see Sect.~\ref{secKM:4}), and using (\ref{eqKM53}),
the differential force in the experimental configuration takes the form
\begin{eqnarray}
&&F_{aan,\rm Au}^{SP}(a)-F_{aan,\rm Si}^{SP}(a)=
-\frac{\pi \hbar^4}{2m_am^2m_{\rm H}^2c^2}(C_{\rm Au}-C_{\rm Si})
\int_1^{\infty}\!\!\!du\frac{\sqrt{u^2-1}}{u^3}
\nonumber \\
&&~~~~~~\times
e^{-2m_acu(a+\widetilde{\Delta}_2+\Delta_1)/\hbar}
\left(1-e^{-2m_acuD/\hbar}\right)
\,X\left(\frac{m_acu}{\hbar}\right),
\label{eqKM55}
\end{eqnarray}
\noindent
where the following notation is introduced
\begin{eqnarray}
X(z)&=&C_{\rm Au}\left[\chi(R,z)-e^{-2z\Delta_2}\chi(R-\Delta_2,z)\right]
\nonumber \\
&+&C_{\rm Cr}e^{-2z\Delta_2}
\left[\chi(R-\Delta_2,z)-e^{-2z\Delta_1}\chi(R-\Delta_2-\Delta_1,z)\right]
\nonumber \\
&+&C_{s}e^{-2z(\Delta_2+\Delta_1)}
\chi(R-\Delta_2-\Delta_1,z)
\label{eqKM56}
\end{eqnarray}
\noindent
and the values of all coefficients $C$ for Au, Cr, Si, and sapphire can be calculated
by using (\ref{eqKM52}) and numerical data for all involved quantities presented in
\cite{KM82}.

The constraints on the parameters of hypothetical forces due to two-axion exchange
between nucleons follow from the inequality
\begin{equation}
|F_{aan,\rm Au}^{SP}(a)-F_{aan,\rm Si}^{SP}(a)|\leqslant\Xi(a),
\label{eqKM57}
\end{equation}
\noindent
where $\Xi(a)$ is the setup sensitivity to force differences in the experiment \cite{KM95}.
This inequality is similar to (\ref{eqKM43}) used in constraining the interaction of
Yukawa type. The strongest current
constraints on the coupling constant of axions to nucleons $g$
follow from (\ref{eqKM57}) in the region of axion masses $4.9~\mbox{meV}<m_ac^2<0.5~$eV.

\begin{figure}[t]
\sidecaption
\includegraphics[scale=.35]{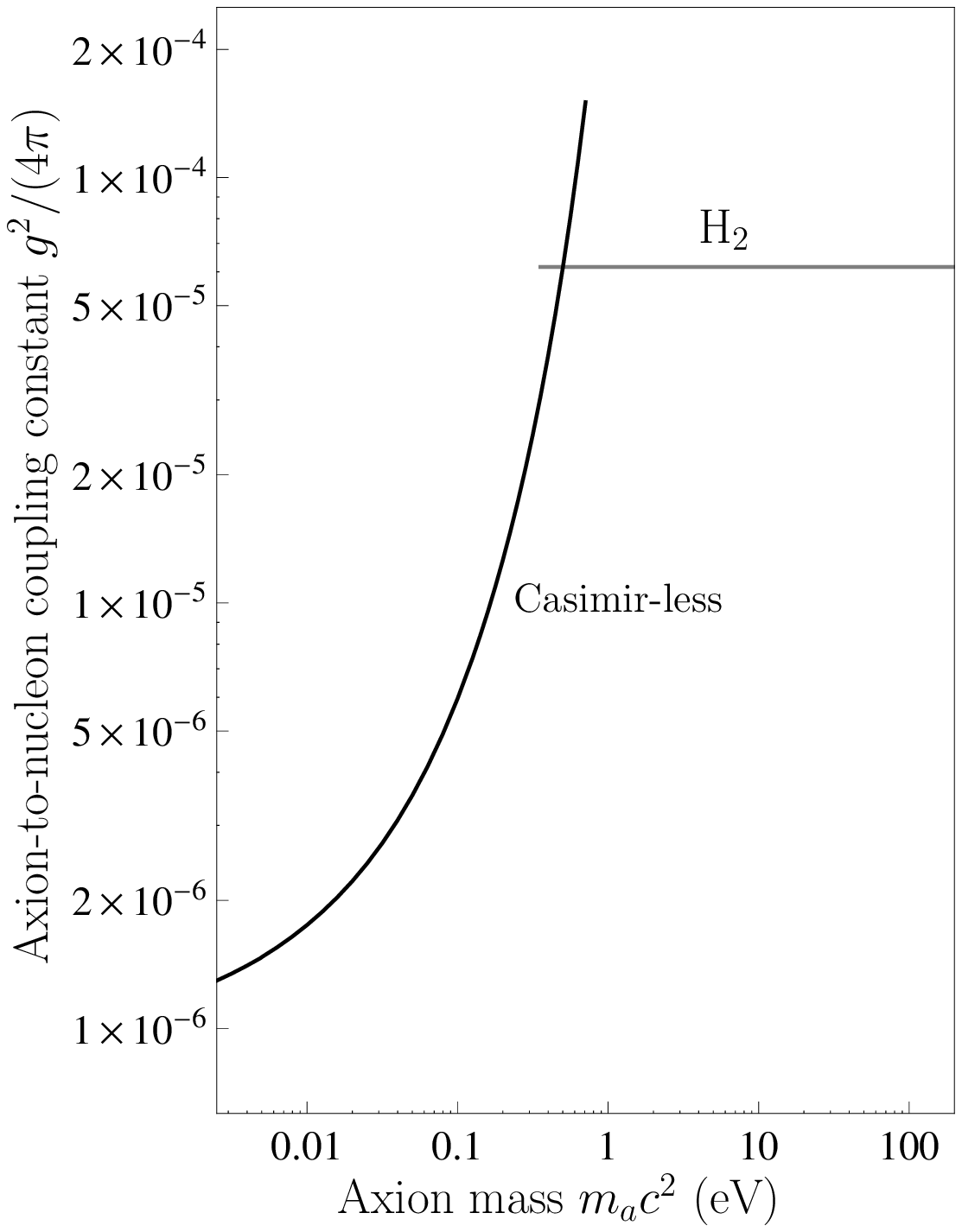}
\caption{Constraints on the coupling constant $g^2/(4\pi)$ of
axions to nucleons are shown as functions of the axion mass $m_ac^2$ by the lines
labeled Casimir-less and H${}_2$ obtained
from the Casimir-less experiment and from measuring dipole-dipole forces between
protons in the beam of molecular hydrogen,  respectively.
The regions of $[m_ac^2,g^2/(4\pi)]$ plane above each line are excluded and below are
allowed}
\label{figKM:4}
\end{figure}
In Fig.~\ref{figKM:4} the obtained constraints are shown by the line labeled Casimir-less.
Similar to all previous figures, the values of axion parameters belonging to the area of
$[m_ac^2,g^2/(4\pi)]$ plane above the line are excluded by the results of differential force
measurements whereas the plane area below the line is allowed.

For $m_ac^2>0.5~$eV the strongest current constraints on $g$ are obtained by comparing
with theory the measurement results for the dipole-dipole forces between two
protons in the beam of molecular hydrogen \cite{KM140,KM141}.
They are shown by the line labeled H${}_2$ in Fig.~\ref{figKM:4}. In this experiment,
the additional force between protons arises due to an exchange of one axion and is described
by the spin-dependent interaction energy (\ref{eqKM47}). As a result, for sufficiently
large $m_a$ the obtained constraints on $g$ are much stronger than those found from the
differential force measurements. What is more, the constraints of line H${}_2$ are valid
both for the originally introduced axions whose interaction with nucleons is described by
the Lagrangian density (\ref{eqKM45}) and for axionlike particles with respective
Lagrangian density (\ref{eqKM46}).

\begin{figure}[t]
\sidecaption
\includegraphics[scale=.35]{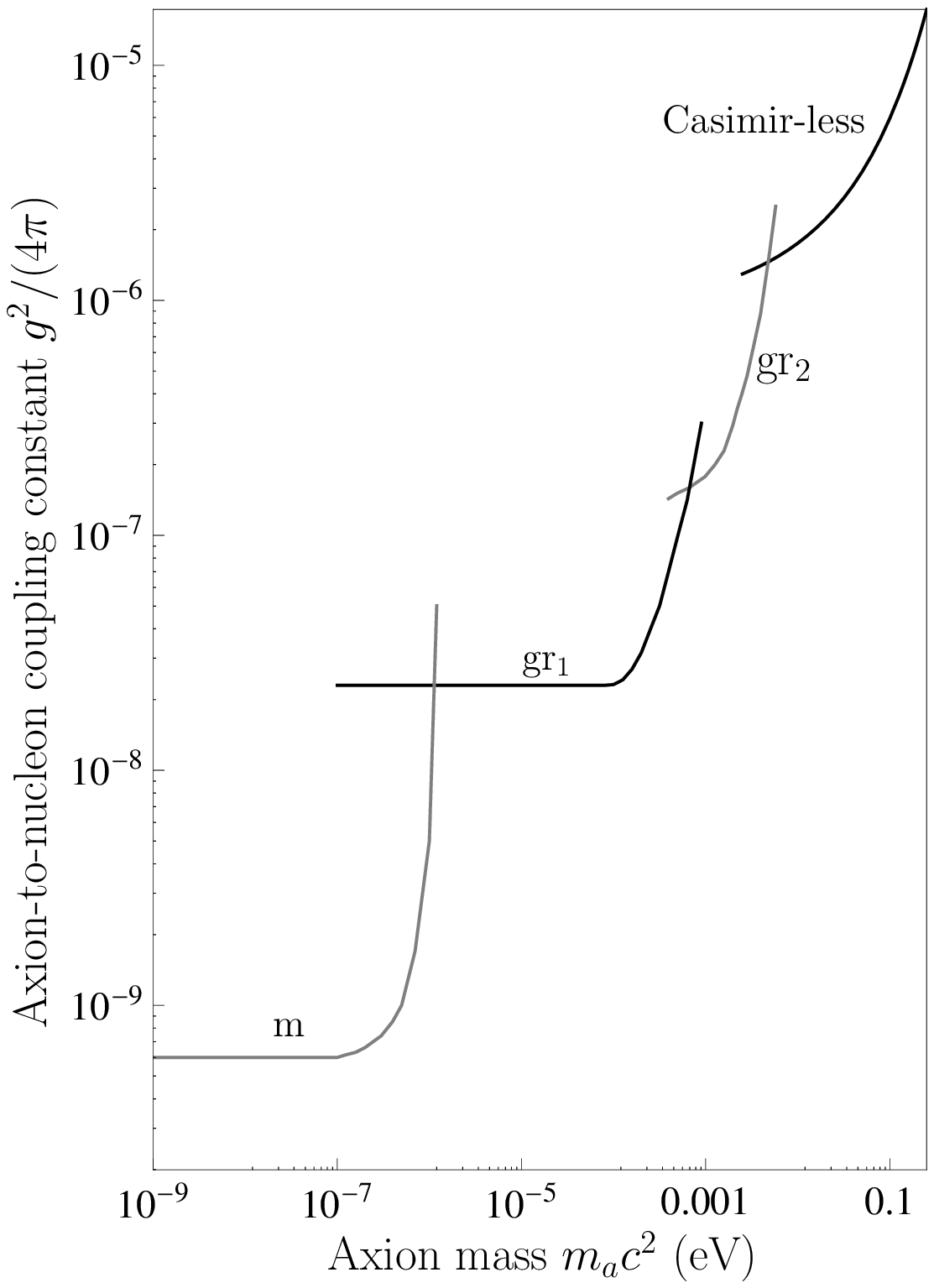}
\caption{Constraints on the coupling constant $g^2/(4\pi)$ of
axions to nucleons are shown as functions of the axion mass $m_ac^2$ by the lines
labeled m, gr${}_1$, gr${}_2$, and Casimir-less obtained from the magnetometer
measurements, Cavendish-type experiment, measuring the minimum force of gravitational
strength, and from the Casimir-less experiment,  respectively.
The regions of $[m_ac^2,g^2/(4\pi)]$ plane above each line are excluded and below are
allowed}
\label{figKM:5}
\end{figure}
In the region of axion masses $m_ac^2<4.9~$meV the strongest constraints of $g$ follow
from gravitational experiments. In Fig.~\ref{figKM:5}, the line labeled gr${}_1$
shows the constraints on $g$ found \cite{KM70} from the Cavendish-type experiment
\cite{KM69}. These constraints are the strongest ones in the region of axion masses
$1~\mu\mbox{eV}<m_ac^2<0.676~$meV. Within the relatively short range of axion masses
$0.676~\mbox{meV}<m_ac^2<4.9~$meV the strongest constraints were obtained \cite{KM137}
by using the planar torsional oscillator for measuring the minimum force of gravitational
strength \cite{KM142,KM143}. The respective constraints are shown by the line labeled
gr${}_2$ in Fig.~\ref{figKM:5}. The gravitational constraints shown by the lines
gr${}_1$ and gr${}_2$ are valid for only the axionlike particles whose interaction
with nucleons is described by the  Lagrangian density (\ref{eqKM46}) because in the
gravitational experiments the test bodies used are unpolarized.

For the smallest axion masses $m_ac^2<1~\mu$eV, the strongest constraints on $g$ were
again obtained from considering the spin-dependent forces which arise due to a one-axion
exchange in the comagnetometer measurements using the spin-polarized K and ${}^3$He
atoms and the  ${}^3$He spin source \cite{KM144}.
These constraints are shown by the line labeled m in Fig.~\ref{figKM:5}.
They are valid for all types of axions and axionlike particles.

The competitive constraints on the parameters of axionlike particles were obtained also
from several other experiments on measuring the Casimir interaction (see, e.g.,
\cite{KM102,KM103,KM145,KM146,KM147,KM148,KM149,KM150}). They are, however, weaker
than those shown by the line labeled Casimir-less and H${}_2$  in Fig.~\ref{figKM:4}.

\section{Could the Casimir Effect be Used For Testing Spin-Dependent Interactions?}
\label{secKM:7}

As explained in the previous section, the process of one-axion exchange between two
nucleons described by the pseudovector Lagrangian density (\ref{eqKM45}) results in the
spin-dependent interaction energy (\ref{eqKM47}) which does not lead to any additional
force between two unpolarized test bodies. The parameters of originally introduced axions
described by this interaction energy were constrained using, e.g., the magnetometer
measurements or from measuring dipole-dipole forces between protons (see Sect.~\ref{secKM:6}).

There are also other predictions of spin-dependent interactions beyond the Standard Model.
In fact the coupling constant of axions to nucleons considered above describes either
the pseudoscalar or pseudovector interactions. This can be notated as $g\equiv g_P$.
In addition to the one-axion exchange, it is possible to consider an exchange of one light
vector particle between two nucleons with a vector and axial vector couplings.
This corresponds to the following Lagrangian density:
\begin{equation}
{\cal L}_{VA}(x)=\hbar c\bar{\psi}(x)\gamma^{\mu}(g_V+g_A\gamma_5)\psi(x)A_{\mu}(x).
\label{eqKM58}
\end{equation}

This Lagrangian density results in the effective spin-dependent interaction energies
between two nucleons \cite{KM144}
\begin{equation}
V_1(r)=\frac{g_A^2}{4\pi r}\hbar
(\mbox{\boldmath$\sigma$}_1\cdot\mbox{\boldmath$\sigma$}_2)e^{-m_Acr/\hbar}
\label{eqKM59}
\end{equation}
\noindent
or
\begin{equation}
V_2(r)=-\frac{g_Ag_V}{4\pi m}\hbar^2
([\mbox{\boldmath$\sigma$}_1\times\mbox{\boldmath$\sigma$}_2]\cdot\mbox{\boldmath$n$})
\left(\frac{m_Ac}{\hbar r}+\frac{1}{r^2}\right)\,e^{-m_Acr/\hbar},
\label{eqKM60}
\end{equation}
\noindent
where $m_A$ is the mass of a vector field $A_{\mu}$. The parameters of the interaction
energies (\ref{eqKM59}) and (\ref{eqKM60}) were constrained by the same comagnetometer
measurements \cite{KM144} which have already been used in Sect.~\ref{secKM:6} for
constraining the interaction energy (\ref{eqKM47}).

Below we discuss the possibility of constraining the spin-dependent
interaction energy (\ref{eqKM47}) from measuring the effective Casimir pressure.
For this purpose it was proposed \cite{KM151} to use the Casimir plates made of silicon
carbide (SiC) with aligned nuclear spins.
It has been known that the nuclear spin of ${}^{29}$Si is equal to 1/2 owing to the
presence of one neutron with an uncompensated spin. In native Si there is only 4.68\%
of the isotope ${}^{29}$Si. In nanotechnology, however, the special procedures are
elaborated for growing the isotopically controlled bulk Si \cite{KM152}.

In \cite{KM151} it was assumed that a fraction of Si atoms $\kappa$ in both plates is
polarized in some definite direction due to the polarization of their nuclear spins
(it was shown that an additional force due to the electronic polarization does not
permit to obtain competitive constraints on the coupling constant of axions to electrons).
In order to obtain the nonzero additional force between plates due to one-axion exchange,
the atomic polarization should be perpendicular to the plates and directed either in one
direction or in the opposite directions \cite{KM151}.

Under these conditions, by integrating the interaction energy (\ref{eqKM47}) over the
volumes of two parallel plates of density $\rho$ and thickness $D$, for the force per
unit area of the plates (i.e., pressure) one obtains \cite{KM151}
\begin{equation}
P_{an}(a)=\pm g^2\frac{\kappa^2\rho^2\hbar^3}{8m^2m_{\rm H}^2c}e^{-m_aca/\hbar}
\left(1-e^{-m_acD/\hbar}\right)^2.
\label{eqKM61}
\end{equation}

The force (\ref{eqKM61}) could be constrained from the experiments \cite{KM27,KM28}
on measuring the effective Casimir pressure using a micromechanical torsional oscillator
if the test bodies were made of SiC with aligned nuclear spins. Similar to (\ref{eqKM41}),
in this case the constraints are obtained from the inequality
\begin{equation}
|P_{an}(a)|\leqslant\Delta P(a).
\label{eqKM62}
\end{equation}

For the pressure (\ref{eqKM61}), the strongest constraints follow at $a=300~$nm, where
$\Delta P(a)=0.22~$mPa \cite{KM27,KM28}, under the conditions that $\kappa=1$ and
$D\gg\hbar/(m_ac)$, i.e., the plates are sufficiently thick. The density of SiC is
$\rho=3.21~\mbox{g/cm}^3$. The most strong constraint that can be placed in this way
on axions and axionlike particles with $m_a=0.0126~$eV is
$g^2/(4\pi)\leqslant 4.43\times 10^{-5}$ \cite{KM151}, which is weaker than that found
from the Casimir-less experiment (see Fig.~\ref{figKM:4}) but is applicable to all
kinds of axions.

An experiment on measuring the Casimir pressure between parallel plates with aligned
nuclear spins can be also used for constraining the interaction of axions with nucleons
from a simultaneous account of one- and two-axion exchange. In this case the constraints
are obtained  from the inequality
\begin{equation}
|P_{an}(a)+P_{aan}(a)|\leqslant\Delta P(a).
\label{eqKM63}
\end{equation}
\noindent
Here the additional pressure between two thick plates due to one-axion exchange is
given by (\ref{eqKM61}), where the last factor on the right-hand side is replaced
with unity, and $P_{aan}$ is obtained by integration of the interaction energy
(\ref{eqKM48}) over the volumes of both plates
\begin{equation}
P_{aan}(a)=-\frac{C_{\rm SiC}^2\hbar^3}{2m^2m_{\rm H}^2c}\int_1^{\infty}\!\!du
\frac{\sqrt{u^2-1}}{u^2}\,e^{-2m_acau/\hbar}.
\label{eqKM64}
\end{equation}
\noindent
The constant $C_{\rm SiC}$ is defined as in (\ref{eqKM52}) using the numerical data
presented in \cite{KM82}.

It has been shown \cite{KM151} that using (\ref{eqKM63}) in the region of axion masses
below 1~eV one could obtain up to an order of magnitude stronger constraints on the coupling
of axions with nucleons than from (\ref{eqKM62}). These constraints, however, would be
valid only for the GUT axions which interaction with nucleons is described by the
pseudoscalar Lagrangian density (\ref{eqKM46}).

\section{Constraining Dark Energy Particles from the Casimir Effect}
\label{secKM:8}

Axions  considered above as the most probable constituents of
dark matter are the Nambu-Goldstone bosons, which appear in the formalism of
quantum field theory when some symmetry (in this case the Peccei-Quinn symmetry) is
broken both spontaneously (i.e., the vacuum is not invariant) and dynamically (i.e., in
the Lagrangian). Although these particles are not the part of the Standarn Model, they
can be considered as its natural supplement. The axionlike particles introduced
later also fall into the standard pattern of quantum field theory.

The particles proposed as the possible constituents of dark energy (chameleons,
symmetrons, etc., see Sect.~\ref{secKM:5}) are quite different.  Unlike all conventional
elementary particles, the properties of these particles depend on the environmental
conditions.

We begin with chameleons whose mass is larger, i.e., the interaction range is shorter,
in environments with higher energy density  (see Sect.~\ref{secKM:5}).
Mathematically chameleons are described by the real self-interacting scalar field
$\Phi$ possessing a variable mass. In the static case, this field satisfies the simplest
equation of the form \cite{KM126,KM153}:
\begin{equation}
\Delta\Phi=\frac{1}{(\hbar c)^4}\frac{\partial V(\Phi)}{\partial\Phi}+
\frac{\rho}{M}e^{\hbar\Phi/(Mc)},
\label{eqKM65}
\end{equation}
\noindent
where $M$ is the typical mass of conventional particles forming the background matter of
density $\rho$ and $V(\Phi)$ is the self-interaction which decreases monotonically with
increasing $\Phi$.

An interaction between chameleons and background matter with density $\rho$ in
(\ref{eqKM65}) implies that the effective interaction potential describing the chameleon
field is given by
\begin{equation}
V_{\rm eff}(\Phi)= V(\Phi)+\rho\hbar^3c^5e^{\hbar\Phi/(Mc)}.
\label{eqKM66}
\end{equation}
\noindent
Although the self-interaction $V$ is assumed to be monotonic, this effective potential
takes the minimum value for $\Phi_0$ satisfying the condition
\begin{equation}
\frac{\partial V(\Phi_0)}{\partial\Phi}+
\frac{\rho(\hbar c)^4}{M}e^{\hbar\Phi_0/(Mc)}=0.
\label{eqKM67}
\end{equation}

Then the mass of the field $\Phi_0$ is given by
\begin{equation}
m^2_{\Phi_0}\equiv\frac{1}{\hbar^2c^6}\frac{\partial^2 V_{\rm eff}(\Phi_0)}{\partial\Phi^2}
=\frac{1}{\hbar^2c^6}\frac{\partial^2 V(\Phi_0)}{\partial\Phi^2}+
\frac{\rho}{M^2}\left(\frac{\hbar}{c}\right)^3e^{\hbar\Phi_0/(Mc)}
\label{eqKM68}
\end{equation}
\noindent
and depends on the background mass density $\rho$.

According to the above assumption, $V$ is a decreasing function of $\Phi$. Then,
$\partial V/\partial\Phi$ is negative and monotonously increasing whereas
$\partial^2 V/\partial\Phi^2$ is positive and decreasing. According to (\ref{eqKM68}),
this means that we have larger $m_{\Phi_0}$ and smaller $\Phi_0$ for larger values of
the background mass density $\rho$ \cite{KM126}.

There are different possible forms of the chameleon self-interaction suggested in the
literature \cite{KM126,KM153,KM154}, e.g.,
\begin{equation}
V(\Phi)=E^4\left(\frac{E}{\hbar c\Phi}\right)^n, \qquad
V(\Phi)=E_0^4\,e^{E^n/(\hbar c\Phi)^n}
\label{eqKM69}
\end{equation}
\noindent
or
\begin{equation}
V(\Phi)=E_0^4\left[1+\left(\frac{E}{\hbar c\Phi}\right)^n\right],
\label{eqKM70}
\end{equation}
\noindent
where $E_0$ and $E$ are the quantities with a dimension of energy and $n=1$, 2, 3, etc.
According to \cite{KM153}, if the chameleon field is responsible for the presently observed
acceleration of the Universe, it should be $E_0\approx 2.4\times 10^{-12}~$GeV.

Similar to axions, an exchange of chameleons between two constituent particles of two
closely spaced test bodies results in some additional force. The constraints on this
force can be obtained from experiments on measuring the Casimir force \cite{KM153}.
In this case, however, both the additional force and constraints on it strongly depend
not only on the specific experimental setup but also on the form of chameleon self-interaction
and other related parameters (see \cite{KM153,KM154,KM155} for some specific results
obtained in the configuration of two parallel plates and a sphere above a plate with
different models of self-interaction).

Another hypothetical particle mentioned in Sect.~\ref{secKM:5} as a possible constituent
of dark energy is a symmetron whose interaction with usual matter becomes weaker
with increasing
mass density of the environment \cite{KM127,KM128}. In the static case the symmetron field
$\Phi_s$ satisfies the equation  \cite{KM127,KM128,KM129}
\begin{equation}
\Delta\Phi_s=\frac{1}{(\hbar c)^4}\frac{\partial V^s(\Phi_s)}{\partial\Phi_s}+
\left[\frac{\hbar}{c}\frac{\rho}{M^2}-\left(\frac{\mu c}{\hbar}\right)^2\right]\Phi_s,
\label{eqKM71}
\end{equation}
\noindent
where $V^s$ is the symmetron self-interaction and $\mu$ is the symmetron mass.

The respective effective potential leading to the right-hand side of (\ref{eqKM71}) is
given by \cite{KM129}
\begin{equation}
V_{\rm eff}^s(\Phi_s)= V^s(\Phi_s)+\frac{1}{2}(\hbar c)^4
\left[\frac{\hbar}{c}\frac{\rho}{M^2}-\left(\frac{\mu c}{\hbar}\right)^2\right]\Phi_s^2,
\label{eqKM72}
\end{equation}
\noindent
where the self-interaction of a symmetron field takes the standard form
\begin{equation}
V^s(\Phi_s)=\frac{1}{4}\lambda\Phi_s^4,
\label{eqKM73}
\end{equation}
\noindent
and $\lambda$ is a dimensionless constant.

The effective potential (\ref{eqKM72}), (\ref{eqKM73}) takes the minimum value for
$\Phi_s=\Phi_{s,0}$ satisfying the condition
\begin{equation}
\lambda\Phi_{s,0}^3+(\hbar c)^4\left[
\frac{\hbar}{c}\frac{\rho}{M^2}-\left(\frac{\mu c}{\hbar}\right)^2\right]\Phi_{s,0}=0.
\label{eqKM74}
\end{equation}

If the density of background matter $\rho<\rho_0=c^3M^2\mu^2/\hbar^3$,
the minimum value of $V_{\rm eff}^s$ is attained at
\begin{equation}
\Phi_{s,0}=\frac{(\hbar c)^2}{\sqrt{\lambda}}\left[
\left(\frac{\mu c}{\hbar}\right)^2-\frac{\hbar}{c}\frac{\rho}{M^2}\right]^{1/2}.
\label{eqKM75}
\end{equation}

Under the opposite condition $\rho>\rho_0$, the minimum value of $V_{\rm eff}^s$ is at
$\Phi_{s,0}=0$. Thus, if $\rho<\rho_0$ the reflection symmetry is broken and  the
vacuum expectation value of $\Phi_{s,0}$ takes a nonzero value. By contrast, in
the regions of high density of background matter $\rho>\rho_0$, the
vacuum expectation value of $\Phi_{s,0}$ turns into zero.

The exchange of symmetrons between two closely spaced material bodies results in some
additional force which was calculated in \cite{KM156} for the experimental configurations
of two parallel plates and a sphere above a plate. According to the results of \cite{KM156},
strong constraints on the parameters of a symmetron can be obtained from measurements of
the Casimir force in these configurations. These measurements, however, are not yet
performed. Prospects in constraining various
hypothetical interactions beyond the Standard Model
and some other laboratory experiments are discussed in the next section.

\section{Outlook}
\label{secKM:9}

Many experiments on measuring the Casimir interaction mentioned above
were performed entirely in an effort to investigate the Casimir effect.
This means that the constraints on corrections to Newtonian gravity
and axionlike particles discussed above were obtained as some kind of
by-product. In \cite{KM157} some improvements in the configurations of
experiments employing both smooth and sinusoidally corrugated surfaces
of a sphere and a plate were suggested which allow obtaining up to an
order of magnitude stronger constraints. Specifically, for the
configurations with corrugated surfaces this could be reached by using
smaller corrugation periods and larger corrugation amplitudes
\cite{KM157}.

There are many proposals of new Casimir experiments aimed for testing
gravity and predictions beyond the Standard Model at short distances.
Thus, it is suggested to measure the Casimir pressure between two
parallel plates at separations up to 10--20 $\mu$m (Casimir and
Non-Newtonian Force Experiment called CANNEX)
\cite{KM158,KM159,KM160,KM161,KM162}.
This experiment promises obtaining  stronger
constraints not only on non-Newtonian gravity and axionlike particles,
but also on chameleon, symmetron and some other theoretical predictions
beyond the Standard Model.

The Casimir-Polder interaction between two atoms or an atom and a cavity
wall can also be used for constraining the hypothetical interactions.
The constraints on an axion to nucleon coupling constant obtained in
this way \cite{KM145} were mentioned in Sect. \ref{secKM:6}. In \cite{KM163} it was
suggested to measure the Casimir-Polder force between a Rb atom and a
movable Si plate screened with an Au film. This makes it possible to
strengthen constraints on the Yukawa interaction constant $\alpha$
in the interaction range around 1 $\mu$m. According to \cite{KM164},
the measured deviations of the Casimir-Polder force between two
polarized particles, arising for photons of nonzero mass, from the
standard one calculated for massless photons can be used for
constraining the extradimensional unification models.

An interesting method for detecting the interaction of axion with
nucleons by means of a levitated optomechanical system was suggested
in \cite{KM165}. In fact this is a version of the Casimir-less
experiment \cite{KM95} where the contribution of the Casimir force
is nullified (see Sects.~\ref{secKM:4} and \ref{secKM:6}). The suggested
method could further
strengthen the already obtained constraints on the coupling constant
of axions to nucleons and on the Yukawa-type corrections to Newtonian
gravity.

There are also many proposed laboratory experiments which are not
closely related to the Casimir physics but could lead to constraints
on the hypothetical interactions in the same or neighboring regions
of parameters as the Casimir effect. Some of them are discussed below.

Thus, the neutron interferometry already used for constraining the
Yukawa-type forces (see Sect.~\ref{secKM:4})  has large potential for
improving the
obtained constraints. Several experiments of this kind have been
performed and suggested pursuing this goal (see, e.g.,
\cite{KM166,KM167,KM168,KM169,KM170}).

There is a continuing interest in the literature to constraining the
power-type, Yukawa-type and other hypothetical interactions by means
of atomic and molecular spectroscopy. A few experiments of this kind
await for their realization \cite{KM171,KM172,KM173}.

It has been known that the levitated nanoparticle sensors are
sensitive to the static forces down to $10^{-17}$ N. In \cite{KM174}
it was suggested to use such sensors for obtaining constraints on the
Yukawa-type corrections to Newtonian gravity. The optomechanical
methods exploiting the levitated sensors were proposed also for
constraining the hypothetical interaction of Yukawa type \cite{KM175}.

Recent literature also contains information on already performed
experiment constraining the exotic interaction between moving
polarized electrons and unpolarized nucleons by means of a magnetic
force microscope \cite{KM176}, on the general scheme allowing an
extraction of constraints on any specific model from different
experiments \cite{KM177}, and on a compressed ultrafast photography
system using temporal lensing for probing short-range gravity
\cite{KM178}.

Interest in all these topics has quickened in the past few years.
One may expect that measurements of the Casimir force and related
table-top laboratory experiments will furnish insights into the
nature of some theoretical predictions beyond the Standard Model
and their relationship to reality.

\vspace*{0.5cm}
\noindent
{\bf Acknowledgments} This is a preprint of the following chapter: Galina L. Klimchitskaya
and Vladimir M. Mostepanenko, Testing Gravity and Predictions Beyond the
Standard Model at Short Distances: The Casimir Effect, published in
Modified and Quantum Gravity, From Theory to Experimental Searches on
All Scales, edited by Christian Pfeifer, Claus Lämmerzahl, 2023, Springer,
reproduced with permission of Springer. The final authenticated version is
available online at: https://doi.org/10.1007/978-3-031-31520-6.
G.L.K. was partially funded by the Ministry of Science and Higher
Education of Russian Federation ("The World-Class Research Center: Advanced
Digital Technologies," contract No. 075-15-2022-311 dated April 20, 2022).
The research of V.M.M. was partially carried out in accordance with the
Strategic Academic Leadership Program "Priority 2030" of the Kazan Federal
University.


\end{document}